\documentclass[12pt,twoside]{article}
\input{epsf.sty}

\def\mypagenumber{1}
\def\mysecnumber{0}  
\def\mydate{April 30, 1998}

\def\mytitle{Bosonized Schwinger model}
\def\myend{\end{document}}

\normalsize
\let\oldtheequation=\theequation
\def\doteqs#1{\setcounter{equation}{0}
            \def\theequation{{#1}.\oldtheequation}}

\newcounter{sxn}
\def\sx#1{\addtocounter{sxn}{1} \bigskip\medskip \goodbreak
\noindent{\bf\leftline{\thesxn.~~#1}} \nobreak \vskip -8pt}
\def\sxn#1{\sx{#1} \doteqs{\thesxn}}

\newcounter{axn}
\def\ax#1{\addtocounter{axn}{1} \bigskip\medskip\goodbreak \noindent{\large\bf
{\Alph{axn}.~~#1}} \nobreak \medskip}
\def\axn#1{\ax{#1} \doteqs{\Alph{axn}}}

\date{}

\newdimen\mybaselineskip
\newcommand{\beeq}{\begin{equation}}
\newcommand{\eneq}{\end{equation}}
\newcommand{\beqn}{\begin{eqnarray}}
\newcommand{\eeqn}{\end{eqnarray}}

\def\ignore#1{{}}

\def\mybig{\displaystyle \strut }

\def\dd{\partial}
\def\la{\raise.16ex\hbox{$\langle$} \, }
\def\ra{\, \raise.16ex\hbox{$\rangle$} }
\def\lla{\la\kern-.4em\la}
\def\rra{\ra\kern-.4em\ra}
\def\go{\rightarrow}

\def\next{{~~~,~~~}}

\def\onehalf{ \hbox{${1\over 2}$} }

\def\onethird{ \hbox{${1\over 3}$} }

\def\psibar{ \psi \kern-.65em\raise.6em\hbox{$-$} }
\def\psibaraL{ \psi^a_L \kern-1.2em\raise.6em\hbox{$-$} ~}
\def\psibaraR{ \psi^a_R \kern-1.2em\raise.6em\hbox{$-$} ~}
\def\mbar{ m \kern-.78em\raise.4em\hbox{$-$}\lower.4em\hbox{} }

\def\Bbar{{\overline B}}
\def\tB{{\widetilde B}}

\def\QED2{QED$_2$}

\def\ep{\epsilon}

\def\vphi{ {\varphi} }

\def\EM{{\rm EM}}
\def\tot{{\rm tot}}
\def\eff{{\rm eff}}

\def\mass{{\rm mass}}
\def\vac{{\rm vac}}
\def\free{{\rm free}}

\def\Wil{\Theta_W}
\def\tf{\tilde f}
\def\hf{\hat f}
\def\tq{\tilde q}
\def\tp{\tilde p}

\def\sign{\,{\rm sign}\,}

\def\Lap{{\triangle}}
\def\LapN{{\triangle_N^\vphi}}

\def\myfrac#1#2{{\mybig #1\over \mybig #2}}

\def\n@space{\nulldelimiterspace=0pt \mathsurround=0pt }
\def\huge#1{{\hbox{$\left#1\vbox to 20.5pt{}\right.\n@space$}}}

\def\boxit#1{$\vcenter{\hrule\hbox{\vrule\kern3pt
    \vbox{\kern3pt\hbox{#1}\kern3pt}\kern3pt\vrule}\hrule}$}
\def\bigbox#1{$\vcenter{\hrule\hbox{\vrule\kern5pt
     \vbox{\kern5pt\hbox{#1}\kern5pt}\kern5pt\vrule}\hrule}$}

\def\myskip{\noalign{\kern 10pt}}
\def\myeqspace{\noalign{\kern 10pt}}
\def\crn{\cr\myskip}

\begin{document}

\footskip 1.0cm

\pagestyle{myheadings}
\markboth{\mytitle}{}

\setcounter{page}{\mypagenumber}

{\baselineskip=15pt \parindent=0pt \small
\mydate 
\hfill   UMN-TH-1610/98
}

\vspace*{40mm}

\centerline{\Large\bf Bosonized Massive N-flavor Schwinger Model}
\vspace*{25mm}
\centerline{\large Yutaka Hosotani$^1$ and Ramon Rodriguez$^{1,2}$}
\vskip .5cm

\centerline{\small $^1$\it School of Physics and Astronomy, University of
Minnesota}
\centerline{\small \it Minneapolis, MN 55455-0112, USA}
\vskip .2cm 
\centerline{\small $^2$\it SDRC, Minneapolis, MN 55425-1528, USA}

\vspace*{25mm}
\parindent=12mm

\centerline{\bf Abstract}
\vskip .5cm 

\centerline{\hbox{\vtop{\hsize=13cm \noindent \small
The massive N-flavor Schwinger model is analyzed by the bosonization method.
The problem is reduced to the quantum mechanics of N degrees of freedom
in which the potential needs to be self-consistently determined by its
ground-state wave function and spectrum with given values of the $\theta$
parameter, fermion masses, and temperature.  Boson masses and fermion
chiral condensates are evaluated.  In the N=1 model the anomalous
behavior is found at $\theta \sim \pi$ and $m/\mu \sim 0.4$.  In the
N=3 model an asymmetry in fermion masses  $(m_1 < m_2 \ll m_3)$ removes the
singularity at $\theta=\pi$ and $T=0$.  The chiral condensates at $\theta=0$
are insensitive to the asymmetry in   fermion masses, but  are
significantly sensitive at $\theta=\pi$.  The resultant picture is 
similar to that obtained in QCD by the chiral Lagrangian method. 
}}}



\newpage

\normalsize
\baselineskip=20pt plus 1pt minus 1pt
\parindent=25pt

\sxn{Introduction}

Two-dimensional quantum electrodynamics (\QED2) 
\cite{SchwGEN}-\cite{Coleman1} with massive N-flavor
fermions resembles
with  four-dimensional quantum chromodynamics (QCD) in many
respects.\cite{HRHI} In both theories non-vanishing chiral condensates  are
dynamically generated.  Fractionally charged test particles are confined in
\QED2, whereas  quarks, or colored objects, are  confined in QCD.
The dynamical chiral symmetry breaking and confinement are not independent
phenomena in \QED2, however.  There would be no confinement if there
were no chiral condensates.\cite{Ellis}-\cite{Ramon}

It has also been recognized that \QED2 describes 
spin systems in nature.\cite{Wiegmann}
  A spin $\onehalf$ anti-ferromagnetic spin chain
is equivalent to two-flavor massless Schwinger model in a uniform background
charge density.  A $n$-leg spin ladder 
systems is equivalent to a coupled set of $n$ Schwinger models.
This equivalence has been successfully employed to account for the 
gap generation in spin ladder systems.\cite{Hosotani_spin}

\QED2 has been analyzed by various methods. On the analytic side
it has been investigated in the perturbation theory, 
in the path integral method, 
and in the bosonization method.
In a series of papers we have shown how to evaluate chiral
condensates and boson mass spectrum for arbitrary values of the $\theta$
parameter, fermion masses ($m$),
and temperature ($T$) by bosonization.\cite{Ramon, HHI, HHI2,massive} 
The mass perturbation theory has been  formulated in the one-flavor 
model.\cite{Adam1,Adam2}   

Investigation in the light-cone quantization method has been pushed
forward both on the analytic and numerical 
sides.\cite{Lenz}-\cite{Harada2}  The bound state spectrum
has been evaluated in the entire range of a fermion mass at
$\theta=0$ and $T=0$. Subtleties
in the chiral condensate in this formalism has been noted.\cite{McCartor}

There has emerged a renewed interest in \QED2 in the lattice gauge theory
approach as well.\cite{Carson}-\cite{deForcrand}  
Recently extensive simulations have been carried out
for the $N=1$ and $N=2$ models. Chiral condensates 
in the $N=1$ model
were evaluated at $\theta=0$ and $T=0$ up to $m/e < 1$.  The boson
mass in the $N=2$ model was evaluated for $m/e < .5$.  After subtracting
condensates in free theory ($e=0$), which depends on regularization methods
employed,  one finds a modest agreement between the bosonization and
lattice results.\cite{deForcrand}

In this paper we further exploit the bosonization method  to investigate
the dependence of chiral condensates and boson mass spectrum on  the
$\theta$ parameter, fermion masses, and temperature.  The advantage of our
method lies in the ability of evaluating physical quantities for arbitrary
values of the $\theta$ parameter and temperature.  The current method,
however, involves an approximation which is not valid for large fermion
masses. Improvement is necessary in this direction.

The Lagrangian of the model is given by
\beeq
{\cal L} = - {1\over 4} \, F_{\mu\nu} F^{\mu\nu} + 
\sum_a \Big\{ \psibar_a \gamma^\mu (i \dd_\mu - e_a A_\mu) \psi_a
- m_a^{} \psibaraR \psi^a_L  - m_a^* \psibaraL \psi^a_R \Big\}
\label{Lagrangian1}
\eneq
where $\psi^a_L = \onehalf (1-\gamma^5)\psi_a$  and
$\psi^a_R = \onehalf (1+\gamma^5)\psi_a$.
Each field carries a charge $e_a$ and mass $m_a$.  We analyze the model
on a circle ($S^1$) with a circumference $L$.  The boundary conditions 
are specified by 
\beqn
&&A_\mu(t,x+L) = A_\mu(t,x)  \cr
&&\psi_a(t,x+L) = - \, e^{2\pi i\alpha_a} \, \psi_a(t,x) ~. 
\label{BC1}
\eeqn

It is important 
to recognize that from the analysis on $S^1$ one can extract physics at 
finite temperature $T$ defined on a line $R^1$.   In the Matsubara
formalism of finite temperature field theory, boson and fermion fields obey
periodic or anti-periodic boundary condition in the imaginary time axis,
respectively;
\beqn
&&A_\mu(\tau+{1\over T},x) = A_\mu(\tau,x)  \cr
\noalign{\kern 10pt}
&&\psi_a(\tau+{1\over T},x) = -  \, \psi_a(\tau,x) ~. 
\label{BC2}
\eeqn
Hence, if one, in a theory defined on $S^1$ with the boundary conditions 
$\alpha_a=0$ in (\ref{BC1}),  analytically continues $t$ from the real axis to 
the imaginary axis, and then interchanges (or relabels) $it$ and $x$, one
arrives at a theory which is exactly the same as the theory defined on 
$R^1$ at $T=L^{-1}$.
This is a powerful equivalence.  One can evaluate chiral condensates, 
Polyakov loops, and various correlators at $T\not= 0$  with the aid of
this correspondence.

This paper is organized as follows.  In Section 2 the bosonized
Hamiltonian is derived on $S^1$.  In Section 3 the $\theta$ vacuum is
introduced and the equation satisfied by its wave function is derived.
In Section 4 we show how the boson mass spectrum and chiral condensates
are evaluated.  Sections 3 and 4 together form a basis of our formulation.
It defines the generalized Hartree-Fock approximation.  The rest of the 
paper is devoted to applying the generalized Hartree-Fock equation to
various models to evaluate the boson spectrum and chiral condensates.
The case of massless fermions is discussed in Section 5.  A useful
truncated formula is derived in Section 6.  The detailed analysis of the 
massive one-flavor model is given in Section 7.  The multi-flavor model
with degenerate fermion masses is analyzed in Section 8.  The case of 
general masses in the N=2 and N=3 models is investigated in Section 9.
A brief summary is given in Section 10. Three appendices collect
useful formulas. 

\vskip .5cm

\sxn{The bosonized Hamiltonian}

Our basic tool is the bosonization method, with which we shall
reduce the model (\ref{Lagrangian1}) to a quantum mechanical system of
finite degrees of freedom. The bosonization method has been developed in
many body theory,\cite{JordanTomonaga} and 
 in field theories on a line $R^1$ \cite{Mandel,Coleman2,Halpern}. The
method has been elaborated on a manifold $S^1$ in the context of string 
theory.\cite{Eguchi}

The bosonization on $S^1$ is particularly unamabiguous, 
sustaining the  absolute rigor. \QED2 on $S^1$ was first studied by
Nakawaki \cite{Nakawaki} and has been developed by many 
authors.\cite{Wolf}-\cite{Iso} It was simplified in ref.\ \cite{HH}, which
we follow in the sequel.  

In this section
we present a brief  review of the method, applying it to the system
(\ref{Lagrangian1}). Although the essence is well known in the literature,
subtle factors associated with the multi-flavor nature and implementation
of arbitrary  boundary conditions  are worth spelling out. With clever
choice of Klein factors the Hamiltonian of the $N$-flavor Schwinger model is
transformed into a surprisingly simple bosonized form.

We note that the $N$-flavor Schwinger model has been analysed in
refs.\ \cite{Halpern,HHI,HHI2,Affleck}.  The model at finite temperature,
which is equivalent to the model on $S^1$, has been analysed in
ref.\ \cite{Love}. The model on other manifolds also
have been investigated in ref.\ \cite{Joos1}.  The conformal
field theory approach to \QED2 has been proposed by Itoi and Mukaida
\cite{Itoi}, which has many  features in common with our bosonization
approach. Correlators of various physical quantities have been discussed in
ref.\ \cite{Smilga1}.

Bosonization of an arbitrary number of fermions on a circle ($0 < x < L$)
obeying  boundary conditions 
\beeq
\psi_a(t,x+L) 
      = -\, e^{2\pi i \alpha_a } \, \psi_a(t,x) 
   \hskip 1cm  (a=1 \sim N) 
\label{BC3}
\eneq
is first carried out in the interaction picture defined by free massless
fermions: $i\gamma \dd \psi =0$.\cite{Halpern,Coleman1}  We introduce
bosonic variables:
\beqn
&&[q^a_\pm, p^b_\pm] = i \, \delta^{ab}  \next
[a^a_{\pm,n}, a^{b,\dagger}_{\pm,m}] = \delta^{ab} \delta_{nm} ~~~, \crn
&&\hbox{all other commutators} = 0 ~~~,\crn
&&\phi^a_\pm (t,x) = \sum_{n=1}^\infty {1\over \sqrt{4\pi n}} \,
  \Big\{ a^a_{\pm,n} \, e^{- 2\pi in(t \pm x)/L} + {\rm h.c.} \Big\} ~~~.
\label{BoseVariables1}
\eeqn
In terms of these variables $\psi_a^t =(\psi^a_+, \psi^a_-)$  can be
expressed as \cite{Eguchi}
\beqn
\psi^a_\pm(t,x) &=& {1\over \sqrt{L}} \, C^a_\pm \,
 e^{\pm i \{ q^a_\pm + 2\pi p^a_\pm (t \pm x)/L \} }
  :\, e^{\pm i\sqrt{4\pi} \phi^a_\pm (t,x) } \, : \crn
&=& \pm {1\over \sqrt{L}} \,
 e^{\pm i \{ q^a_\pm + 2\pi p^a_\pm (t \pm x)/L \} } \, C^a_\pm 
  :\, e^{\pm i\sqrt{4\pi} \phi^a_\pm (t,x) } \, : 
\label{bosonization1}
\eeqn
where the Klein factors are given by \cite{HHI}
\beqn
C^a_+ &=& \exp \bigg\{ i\pi \sum_{b=1}^{a-1} ( p^b_+ + p^b_- - 2 \alpha_b)
\bigg\}  \crn 
C^a_- &=& \exp \bigg\{ i\pi \sum_{b=1}^{a} ( p^b_+ - p^b_-) \bigg\}  
\label{Klein1} 
\eeqn
This choice of Klein factors  turns out very convenient. They satisfy
\beqn
\left[ \matrix{ C^a_+\cr{C^a_+}^\dagger\cr} \right]  \, \psi^b_\pm(x) 
&=& \sign (b\ge a)  \, \psi^b_\pm(x)   \, 
\left[ \matrix{ C^a_+\cr{C^a_+}^\dagger\cr} \right] \crn
\left[ \matrix{ C^a_-\cr{C^a_-}^\dagger\cr} \right]  \, \psi^b_\pm(x) 
&=& \sign (b > a)  \, \psi^b_\pm(x)  \, 
\left[ \matrix{ C^a_-\cr{C^a_-}^\dagger\cr} \right] 
\label{Klein2}
\eeqn
where  $\sign (A)$ is defined to be $+1$ ($-1$), when $A$ is true
(false).   By construction $(\dd_t \mp \dd_x ) \, \psi^a_\pm (t,x) = 0$, 
i.e.\ $\psi_a$ satisfies a free massless Dirac equation:
$(\gamma^0 \dd_0 + \gamma^1 \dd_1) \, \psi^a =0$ where
$\gamma^\mu = (\sigma_1, i\sigma_2)$ and 
$\gamma^5 = \gamma^0 \gamma^1 = - \sigma_3$.

Under a translation along the circle 
\beeq
 \psi^a_\pm (t,x+L) = - e^{2\pi i p^a_\pm} \, \psi^a_\pm (t,x) =
-  \psi^a_\pm (t,x) \, e^{2\pi i p^a_\pm}    
\label{BC4}
\eneq
so that the boundary  condition (\ref{BC1}) is ensured by a physical state
condition
\beeq
e^{2\pi i p^a_\pm} ~ | \, {\rm phys} \ra = e^{2\pi i \alpha_a} ~ 
|\, {\rm phys} \ra ~~. 
\label{PhysCond1}
\eneq
Further conditions can be consistently imposed on physical states such
that  the Klein factors act in physical space as a semi-identity operator:
$C^a_\pm  ~ | \, {\rm phys} \ra = (+ ~{\rm or}~-)  | \, {\rm phys} \ra$.
We shall see below that the Hamiltonian commutes with $p^a_+ - p^a_-$.

The fields $\{ \psi^a_\pm(x) \}$ satisfies desired equal-time anti-commutation
relations.  With the aid of (\ref{Klein2}), (\ref{formula2}) and
(\ref{formula3}),  it is straightforward to show
\beqn
\{ \psi^a_\alpha(x), \psi^b_\beta(y)^\dagger \} 
&=& \delta^{ab} \delta_{\alpha\beta} ~ e^{i\pi (x-y)/L} \cdot e^{2\pi i
p^a_+(x-y)/L}  ~ \delta_L(x-y)  \cr
\noalign{\kern 6pt}
\hbox{others} &=& 0 
\label{commutation1}
\eeqn
where $\alpha, \beta= +$ or $-$.
Notice that the extra phase factors in (\ref{commutation1}) manifest the
translation property (\ref{BC4}).
The bosonization in the interaction picture is defined by (\ref{bosonization1})
and (\ref{PhysCond1}).

In applying the bosonization method to the model (\ref{Lagrangian1}), 
it is most convenient to take the Coulomb gauge
\beqn
&&A_1(t,x) = b(t)  ~~~, \crn
&&A_0(t,x) = - \int_0^L dy \, G(x-y) \, j_\EM^0(t,y) 
 ~~~,~~~ j_\EM^0 = \sum_a e_a \psi_a^\dagger \psi_a  ~~~, \cr
&&G(x+L) = G(x) ~~~,~~~ {d^2\over dx^2} \, G(x) = \delta_L(x) - {1\over L} 
\label{Coulomb1}
\eeqn
in which the zero mode $b(t)$ of $A_1(t,x)$ is the only physical degree of
freedom associated with the gauge fields.
The Hamiltonian is given by
\beqn
&&H = {1\over 2L} \, P_b^2 +
\int_0^L dx \, \sum_a \Big\{ \psibar_a  \gamma^1 (-i\dd_1 + e_a b)\psi_a
 + m_a^{} \psibar_{aL} \psi_{aR} + m_a^* \psibar_{aR} \psi_{aL} \Big\} \crn
&&\hskip 4cm - {1\over 2} \int_0^L dxdy ~ j_\EM^0(x) G(x-y) j_\EM^0(y) ~~.  
\label{Hamiltonian1}
\eeqn 
Here $P_b$ is the  momentum conjugate to $b$:  $P_b= L \dot b$.  The
anti-symmetrization of fermion operators is understood.

At all stages of bosonization, the gauge invariance must be maintained.
Due caution is necessary in bosonizing a product of two field operators
at the same point as it has to be regularized.  At equal time the 
bosonization formula (\ref{bosonization1}) leads to
\beqn
&&e^{-i e_a b(y-x)} \, {1\over 2}
   [\,\psi^a_\pm (y)^\dagger ~,~ \psi^a_\pm (x)\,]  \cr 
&&= \mp {1\over 2\pi i} \, \bigg\{ {{\rm P} \over x-y} 
 + i \,\Big( {2\pi p^a_\pm \over L} + e_a b 
  \pm  \sqrt{4\pi}  \dd_y \phi^a_\pm \Big)\cr
&&\quad + (x-y) \Big[ {\pi^2\over 6 L^2} - {1\over 2} \, 
: \Big( {2\pi p^a_\pm \over L} + e_a b 
  \pm  \sqrt{4\pi} \dd_y \phi^a_\pm \Big)^2 : 
\pm {i \sqrt{4\pi} \over 2} \, {\dd^2 \phi^a_\pm\over \dd y^2} \Big] 
   + \cdots  \bigg\}   
\label{bosonization2}
\eeqn
Gauge invariant regularization  amounts to  dropping the P$/(x-y)$ term.

Hence we have
\beqn
&&{1\over 2} [\, \psi^{a \dagger}_\pm , \psi^a_\pm \,] =
\mp {1\over 2\pi} \, \Big( {2\pi p^a_\pm\over L} + 
  e_a b \pm  \sqrt{4\pi} \dd_x \phi^a_\pm \Big) \crn
&&\pm {1\over 2} [\, \psi^{a \dagger}_\pm ,(i\dd_x - e_a b) \psi^a_\pm \,]
\crn
&&\hskip 1cm = - {\pi\over 12L^2} + {1\over 4\pi}  
: \Big( {2\pi p^a_\pm \over L} + e_a b 
 \pm  \sqrt{4\pi} \dd_y \phi^a_\pm \Big)^2 : 
 \mp {i\over  \sqrt{4\pi} } \dd^2_x \phi^a_\pm \crn
&&\pm {1\over 2} [\, (-i\dd_x - e_a b) \psi^{a \dagger}_\pm , \psi^a_\pm \,]
\crn
&&\hskip 1cm = - {\pi\over 12L^2} + {1\over 4\pi}  
: \Big( {2\pi p^a_\pm \over L} + e_a b 
  \pm  \sqrt{4\pi} \dd_y \phi^a_\pm \Big)^2 : 
 \pm {i\over \sqrt{4\pi} } \dd^2_x \phi^a_\pm     
\label{bosonization3}
\eeqn  
In particular, currents are given by
\beqn
j^0_a &=& + \onehalf [ \psi^{a \dagger}_+,\psi^a_+] 
  + \onehalf [\psi^{a \dagger}_-, \psi^a_-] =
{-p^a_+ + p^a_- \over L} - {1\over \sqrt{\pi} }  \, \dd_x \phi_a \crn
j^1_a &=& - \onehalf [ \psi^{a \dagger}_+,\psi^a_+] 
  + \onehalf [\psi^{a \dagger}_-, \psi^a_-] =
{+p^a_+ + p^a_- \over L}  + e_a {b \over \pi} 
+ {1\over \sqrt{\pi}} \, \dd_t \phi_a 
\label{current1}
\eeqn
where $\phi_a = \phi^a_++\phi^a_-$.  In terms of $\tilde \phi_a =
\tilde\phi_{a+} + \tilde\phi_{a-}$ where
$\tilde \phi_{a\pm} = (4\pi)^{-1/2} \, [  q^a_\pm
 + 2\pi  p^a_\pm \, (t\pm x)/ L\, ] + \phi^a_\pm$,
the current takes a simpler form 
\beeq
j^\mu_a = - {1\over \sqrt{\pi}} \, \ep^{\mu\nu} \dd_\nu \tilde\phi_a 
 + \delta^{\mu 1} {e_a b\over \pi} ~~.
\label{current2}
\eneq
In the following discussions, however, we shall find that treating the zero
mode  and oscillatory mode parts separately is more convenient.

The kinetic energy term is transformed  to
\beqn
-i \psibar_a \gamma^1 D_1 \psi_a \equiv 
{i\over 4} \, \big\{ [ \psi^{a \dagger}_+,D_1 \psi^a_+ ] 
 - [ D_1 \psi^{a \dagger}_+, \psi^a_+ ] -[ \psi^{a \dagger}_-,D_1 \psi^a_- ] 
+ [ D_1\psi^{a \dagger}_- ,\psi^a_- ] \big\} \hskip 1cm &&\crn
\noalign{\kern 7pt}
=- {\pi\over 6L^2} + 
{1\over 4\pi}  :\Big( {2\pi p^a_+ \over L} + e_a b +
 \sqrt{4\pi}  \dd_x \phi^a_+ \Big)^2 : 
+ {1\over 4\pi} :\Big( {2\pi p^a_- \over L} + e_a b - 
  \sqrt{4\pi} \dd_x \phi^a_- \Big)^2: ~. &&
\label{kinetic-energy}
\eeqn
Putting all things together,  we find
\beqn
H ~ &=& H_0 + H_\phi + H_\mass \crn
H_0 &=&
 - {\pi N\over 6 L} + {P_b^2\over 2L} 
+ {\pi\over 2L} \sum_{a=1}^N \Big\{ (p^a_+-p^a_-)^2 
+ (p^a_+ + p^a_- + {e_a bL\over \pi} )^2 \Big\} \crn
H_\phi &=&
 \int_0^L dx \, {1\over 2} 
: \Big\{
 \sum_{a=1}^N ( \dot\phi_a^2 + \phi_a'^2 )
     + \mu^2 \,  \bar\phi^2 \Big\} : \crn
&&\mu^2 = {1\over \pi} \sum_a e_a^2 \equiv {\bar e^2\over \pi} ~~,~~
\bar\phi = \sum_a {e_a\over \bar e} \phi_a  
\label{Hamiltonian2}
\eeqn
where $H_\mass$ represents the fermion mass term.  It is  convenient
to express the Hamiltonian in terms of
\beqn
&&q_a = q^a_+ + q^a_- \next p_a = \onehalf (p^a_+ + p^a_-) \cr
&&\tilde q_a = \onehalf (q^a_+ - q^a_-) \next \tilde p_a = p^a_+ - p^a_-\cr
&&[q_a, p_b] = [\tilde q_a, \tilde p_b] = i \delta_{ab}
\next \hbox{all others} = 0 ~~~;
\label{new-zeromodes}
\eeqn
$H_0$ becomes
\beeq
H_0 =
 - {\pi N\over 6 L} + {P_b^2\over 2L} 
+ {1\over 2\pi L} \sum_{a=1}^N \Big\{ 
 ( e_a bL + 2\pi p_a)^2 + \pi^2 \tilde p_a^2 \Big\} ~~.
\label{Hamiltonian3}
\eneq

A few important conclusions can be drawn in 
 the massless fermion case ($H_\mass=0$).  The zero mode part $H_0$ and
oscillatory part $H_\phi$ decouple from each other.   Each part is 
bilinear in operators so that the Hamiltonian is  solvable.
The oscillatory part consists of one massive boson ($\bar\phi$) with a mass
$\mu$ and $N-1$ massless bosons.  The zero mode part must be solved with
the physical state condition 
\beqn
&&e^{2\pi i p^a_\pm} ~ | \, {\rm phys} \ra = e^{2\pi i \alpha_a} ~ 
|\, {\rm phys} \ra ~, \crn 
&&Q^\EM  |{\rm phys}\ra = - \sum_a e_a \tilde p^a |{\rm phys}\ra  =0 ~.
\label{PhysCond3}
\eeqn

\vskip .2cm

\sxn{\bf $\theta$ vacuum}

When all ratios of various charges $e_a$ are  rational, there 
results a $\theta$ vacuum.  In this article we restrict ourselves to
the case in which all fermions have the same charges:  $e_a = e$.
It is appropriate to introduce the Wilson line phase $\Wil$:
\beeq 
\Wil = e b L  \next e^{i\Wil (t)} = e^{i e \int_0^L dx \, A_1(t,x)} ~~~.
\label{Wilson1}
\eneq
The zero mode part of the Hamiltonian (\ref{Hamiltonian3}) becomes
\beeq
H_0 =
 - {\pi N\over 6 L} + {\pi \mu^2 L\over 2N} P_W^2
+ {1\over 2\pi L} \sum_{a=1}^N \Big\{ 
 ( \Wil + 2\pi p_a)^2 + \pi^2 \tilde p_a^2 \Big\} 
\label{Hamiltonian4}
\eneq
where $\mu^2= Ne^2/\pi$ and $P_W=\dot \Wil / e^2L$ is the conjugate
momentum to $\Wil$.   In terms of the new variables
\beqn
&&\Wil' = \Wil + {2\pi\over N} \sum p_a \next
  q_a' = q_a + {2\pi\over N} P_W \crn
&&[\Wil' , P_W] = i \next [q_a', p_b] = i \delta_{ab} \next 
{\rm others} = 0 
\label{newWil}
\eeqn
the Hamiltonian becomes
\beeq
H_0  =
 - {\pi N\over 6 L} + {\pi \mu^2 L\over 2N}  P_W^2
+ {N\over 2\pi L}  {\Wil'}^2 
- {2\pi\over NL} \Big( \sum_a p_a \Big)^2 
+ {2\pi\over L} \sum \Big( p_a^2 + {1\over 4}  \tilde p_a^2 \Big) 
~~.
\label{Hamiltonian5}
\eneq
There appears an additional symmetry $(\Wil, p_a) \go (-\Wil, -p_a)$
when $\alpha_a=0$ and $\onehalf$. The Hamiltonian is invariant under
\beeq
\Wil \go \Wil + 2\pi \next p_a \go p_a - 1 
\label{largeGT1}
\eneq
or equivalently, in terms of the original fields,
\beeq
A_\mu \go A_\mu + {1\over e} \dd_\mu \Lambda \next \psi_a \go
e^{i\Lambda} \psi
\next \Lambda = {2\pi  x\over L} ~~.
\label{largeGT2}
\eneq
The transformation is generated by a unitary operator
\beqn
&&U = e^{i(2\pi P_W + \sum q_a)} = e^{i\sum q_a'} \cr
&& [U, H] =0 ~~~.
\label{largeGT3}
\eeqn

In a vector-like theory $\tilde p_a=p^a_+ - p^a_-$ takes integer eigenvalues. 
Further $[\tilde p_a,H]=0$.  We can restrict ourselves to states
with $\tilde p_a =0$  as the energy is minimized there.  The vacuum state is
written as a direct product of ground states of the zero mode sector
and oscillatory ($\phi$) mode sector.   The ground state in the oscillatory
mode  sector is defined with respect to physical boson masses $\mu_\alpha$'s.
As we shall see, the ground state wave function in the zero mode sector affects
the physical boson masses, and vice versa.  These two must be determined 
self-consistently.
Note that if there is a background charge ($Q_{b.g.} \not= 0$)
as in the case of spin chains, then  
$- \sum e_a \tilde p_a = - Q_{b.g.}$.   

With this understanding the vacuum wave function is written as
\beeq
|\Psi_\vac\ra = \int_{-\infty}^\infty dp_W \sum_{\{n,r_a\}}
    |p_W, n, r_a\ra ~  \tf (p_W,n, r_a)
\label{vacuum1}
\eneq
where $|p_W, n, r_a\ra$ is an eigenstate of $P_W$ and $p_a$:
\beqn
 P_W \, | p_W, n, r_a\ra  &=& p_W\,  | p_W, n, r_a\ra \crn
 p_a\,| p_W, n, r_a\ra  
&=& \cases{ ( n+ r_a +\alpha_a  )\, | p_W, n, r_a\ra 
    &for $a<N$\cr
( n +\alpha_N  )\, | p_W, n, r_a\ra   &for $a= N$.\cr} 
\label{state1}
\eeqn
It follows that
\beqn
e^{i k \theta_W} \, | p_W, n, r_b\ra &=&  ~~ | p_W + k, n, r_b\ra \crn
e^{\pm i q_a} \, | p_W, n, r_b\ra &=& \cases{
\, | p_W, n, r_b \pm \delta_{b,a}\ra &for $a<N$\cr
\, | p_W, n\pm 1, r_b \mp 1\ra &for $a= N$.\cr} 
\label{state2}
\eeqn

Since $U$ in (\ref{largeGT3}) commute with the Hamiltonian, one can take
\beeq
U\, |\Psi_\vac (\theta) \ra = e^{+i\theta} \, |\Psi_\vac(\theta) \ra  ~~.
\label{vacuum2}
\eneq
As $U |p_W, n, r_a\ra = e^{2\pi ip_W} |p_W, n+1, r_a\ra$, one finds
$\tf(p_W,n,r_a) 
= e^{-in\theta + 2\pi i (n+\bar\alpha) p_W} \,  \tf(p_W , r_a)$ where
$\bar\alpha = N^{-1} \sum_{a=1}^N \alpha_a$.

This is the $\theta$ vacuum.  The existence of the operator $U$ was
suspected  long time ago in ref.\ \cite{CJS}, though its explicit form was
not given.  On a circle $U$ is unambiguously written in terms of $P_W$
and $q_a$'s.  This definition was first given in ref.\ \cite{HH}. 
In passing, it has been noticed recently that  similar definition  of the
$\theta$ vacuum arises in the framework of light cone quantization of the 
Schwinger model.\cite{Kalloniatis,Martinovic}

It is convenient to adopt a coherent state basis given by
\beeq
|p_W, n, \vphi_a\ra = {1\over (2\pi)^{(N-1)/2}}  \sum_{\{ r_a \}}
e^{- i (r_1\vphi_1 + \cdots + r_{N-1}\vphi_{N-1}) } ~ |p_W, n,  r_a\ra ~.
\eneq
Then
\beqn
|\Psi_\vac(\theta) \ra &=& 
{1\over \sqrt{2\pi}} \sum_n \int dp_W [d\vphi] 
~ |p_W, n, \vphi_a\ra  ~e^{-in\theta + 2\pi i (n+\bar\alpha) p_W} ~ 
   \hf(p_W, \vphi_a) \crn
\hf(p_W, \vphi_a) &=& {1\over (2\pi)^{(N-1)/2}} \sum_{\{ r_a \}}
e^{i\sum r_a\vphi_a} ~ \tf(p_W, r_a)  ~~~.
\label{vacuum4}
\eeqn
The normalization is
$\la \Psi_\vac(\theta') | \Psi_\vac(\theta) \ra = \delta_{2\pi}
(\theta'-\theta)$ and 
$\int dp_W [d\vphi] ~ \big| \hf  \big|^2 =1$. 

The function $\hf(p_W, \vphi_a)$ is determined by solving the eigenvalue
equation
$H_\tot  | \Psi_\vac(\theta) \ra = E_\vac  | \Psi_\vac(\theta) \ra$.
We write, for an operator $Q=Q(\Wil, P_W, p_a)$,
\beeq
Q \, |\Psi_\vac(\theta) \ra 
={1\over \sqrt{2\pi}}\sum_n \int dp_W [d\vphi]  |p_W,n, \vphi_a\ra 
~ e^{-in\theta + 2\pi i (n+\bar\alpha) p_W}  
 \, \hat Q  \, \hf(p_W , \vphi_a) ~.
\label{DefineHat}
\eneq
Noticing 
$\Wil |p_W,n, \vphi_a\ra = -i(\dd/\dd p_W) |p_W,n, \vphi_a\ra $, one finds
 that
\beeq
Q =  \Wil + {2\pi\over N} \sum_{a=1}^N p_a  
\quad \Rightarrow \quad 
\hat Q = i \bigg\{ {\dd\over \dd p_W} - {2\pi\over N}
\sum_{a=1}^{N-1} {\dd\over \dd\vphi_a} \bigg\} ~~.
\label{Schro1}
\eneq
The operator $\hat H_0$ corresponding to $H_0$ is
\beeq
\hat H_0 =  - {\pi N\over 6L} + {e^2L\, p_W^2\over 2}  
- {N\over 2\pi L} \bigg( {\dd\over \dd p_W} 
    - {2\pi \over N} \sum_{a=1}^{N-1} {\dd\over \dd\vphi_a} \bigg)^2 
-  {2\pi (N-1)\over NL} ~ \Lap_\vphi ~~.
\label{Schro3}
\eneq
Here the $\vphi$-Laplacian is given by
\beeq
\Lap_\vphi = \sum_{a=1}^{N-1} 
\Big( {\dd\over \dd \vphi_a} - i \beta_a \Big)^2
- {2\over N-1} \sum_{a<b}^{N-1} 
\Big( {\dd\over \dd \vphi_a} - i \beta_a \Big)
\Big( {\dd\over \dd \vphi_b} - i \beta_b \Big) 
\label{Laplacian}
\eneq
where $\beta_a = \alpha_a - \alpha_N$.

The mass operator in the Schr\"odinger picture is 
\beeq
M_{aa}^S = \psi^{a\dagger}_- \psi^a_+  
=  - C^{a \dagger}_- C^a_+\cdot e^{ - 2\pi i \tilde p^ax/L }
\, e^{i q^a} ~ L^{-1} \, N_0[ e^{i\sqrt{4\pi}\phi_a} ]  
\label{massOperator1}
\eneq
where the normal ordering $N_0[~~]$ is defined with respect to massless
fields.  In the presence of $H_\mass$, all boson fields
$(\phi)$ become massive.  We denote a mass eigenstate with a mass
$\mu_\alpha$ by $\chi_\alpha$:
\beeq
\chi_\alpha = U_{\alpha a} \, \phi_a \next U^t U = I ~~~.
\label{chi-fields}
\eneq
The vacuum is defined with respect to these $\chi_\alpha$ fields.

With the aid of (\ref{B-function1}), the mass
operator becomes 
\beqn
M_{aa}^S 
&=&  - C^{a \dagger}_- C^a_+\cdot e^{ - 2\pi i \tilde p^ax/L }
\, e^{i q^a} ~  L^{-1} \, \Bbar_a  \prod_{\alpha=1}^N  
  N_{\mu_\alpha} [ e^{iU_{\alpha a}\sqrt{4\pi}\chi_\alpha} ]  \crn
\Bbar_a &=& \prod_{\alpha=1}^N B(\mu_\alpha L)^{(U_{\alpha a})^2}  
\label{massOperator2}
\eeqn
Further
\beqn
&&\la p_W', n', \vphi' | e^{\pm iq_a} | p_W, n, \vphi\ra \crn
\noalign{\kern 5pt}
&&\hskip 0cm = 
\delta(p_W'-p_W) \prod_{b=1}^{N-1} \delta_{2\pi}(\vphi_b'-\vphi_b) 
\cases{
\delta_{n',n} ~ e^{\pm i \vphi_a}  &for $a<N$\cr\cr
\delta_{n',n\pm1} ~ e^{\mp i \sum_b\vphi_b}  &for $a=N$\cr} 
\label{qMatrix1}
\eeqn
so that
\beeq
Q={e^{\pm iq_a}} \Rightarrow 
\hat Q =\cases{ e^{\pm i \vphi_a} &for $a<N$\cr\cr
        e^{\pm i(\theta-\sum\vphi_b -2\pi p_W)} &for $a=N$.\cr}
\label{qMatrix2}
\eneq

Let us write a fermion mass as $m_a= |m_a| \, e^{i\delta_a}$ and
drop the absolute value sign henceforth.  Then
$H_\mass= \int dx \,\sum_a m_a (M_a e^{i\delta_a} + ~{\rm h.c.}~)$.  
When $H_\mass$  acts on$|\Psi_\vac(\theta)\ra$,  in
general $\chi_\alpha$ quanta are also excited as well as in the
zero-mode sector.  In deriving an equation for
$\hf(p_W,\vphi)$, we ignore those $\chi$-excitations, with the
understanding that physical masses $\mu_\alpha$'s are taken.  Then
\beeq
\hat H_\mass =
-  \sum_{a=1}^{N-1} 2 m_a  \Bbar_a
\cos( \vphi_a+\delta_a) - 2 m_N \Bbar_N
 \cos(\theta-\sum\vphi_b -2\pi p_W + \delta_N) ~.
\label{massMatrix1}
\eneq
$\hf(p_W, \vphi_a)$ must satisfy 
$\hat H_\tot \hf(p_W, \vphi_a) = E_\vac \hf(p_W, \vphi_a)$ where 
$\hat H_\tot= \hat H_0 + \hat H_\mass$.

At this stage we recognize that it is appropriate to introduce
\beeq
f(p_W, \vphi_a) = \hf(p_W, \vphi_a- {2\pi p_W\over N} -\delta_a)  ~~~.
\label{effctiveTheta}
\eneq
The eigenvalue equation $\hat H_\tot \hf(p_W, \vphi_a) = E_\vac \hf(p_W,
\vphi_a)$ now reads
\beeq
\Bigg\{
- \bigg( {N\over 2\pi} \bigg)^2 {\dd^2\over \dd p_W^2} 
- (N-1) \Lap_\vphi + V(p_W,\vphi) \Bigg\} f(p_W, \vphi) = \ep
f(p_W,\vphi)  ~~~.
\label{Schro4}
\eneq
where $\ep = (NLE_\vac/ 2\pi) + (\pi N^2/ 12)$.
The potential is given by
\beqn
&&V (p_W,\vphi) = + {(\mu L)^2\over 4}  \, p_W^2  - {N\over \pi}
\sum_{a=1}^N m_a L \Bbar_a 
       \cos \Big( \vphi_a - {2\pi p_W\over N}\Big) \crn
&& 
\vphi_N =\theta_\eff - \sum_{a=1}^{N-1} \vphi_a 
\next \theta_\eff = \theta + \sum_{a=1}^N \delta_a ~.
\label{potentail1}
\eeqn
The vacuum is
\beqn
&&|\Psi_\vac(\theta) \ra = 
{1\over \sqrt{2\pi}} \sum_n \int dp_W [d\vphi] ~ |p_W, n, \vphi_a\ra \crn
&&\hskip 3cm  \times ~ e^{-in\theta + 2\pi i (n+\bar\alpha) p_W} ~ 
   f(p_W, \vphi_a+{2\pi p_W\over N} + \delta_a) ~.
\label{vacuum5}
\eeqn
The problem has been reduced 
to solving the Schr\"odinger equation for $N$ degrees of freedom.

\vskip .5cm

\sxn{Boson masses and condensates}

In deriving the equation (\ref{Schro4}) for the vacuum wave function
$f(p_W,\vphi)$,  we have supposed that we already know the masses
$\mu_\alpha$'s of the boson fields $\chi_\alpha$'s.  In this section
we show how these $\mu_\alpha$'s are related to the vacuum wave function
$f(p_W,\vphi)$ itself.  Hence we obtain a self-consistency condition for
the vacuum.  Along the way we shall also find that the chiral condensate
$\la \psibar_a \psi_a\ra$ is related to $\mu_\alpha$'s.

From (\ref{qMatrix2}) it follows that
\beqn
\la e^{\pm i q_a} \ra_\theta 
&=& \lim_{\theta' \go \theta}  
\la \Psi_\vac(\theta') |  e^{\pm i q_a} | \Psi_\vac(\theta) \ra \Big/
\la \Psi_\vac(\theta') |   \Psi_\vac(\theta) \ra \crn
&=& e^{\mp i \delta_a} \, \la e^{\pm i (\vphi_a-2\pi p_W/N)} \ra_f 
\label{qMatrix3}
\eeqn
where $\vphi_N$ is defined in (\ref{potentail1}) and the $f$-average is
given by $\la F \ra_f = \int dp_W [d\vphi] \,  F |f|^2$.
In our approximation scheme $\la M_{aa}\ra_\theta = 
 - ( \Bbar_a/L) \la e^{iq_a} \ra_\theta$, and therefore
\beeq
\la \psibar_a\psi_a\ra_\theta' 
= - {2\over L} \Bbar_a \la \cos (\vphi_a -{2\pi p_W\over N} - \delta_a)
\ra_f ~~. 
\label{condensate1}
\eneq
Rigorously speaking, the formula (\ref{condensate1}) is valid only
for small $m$.  Adam has determined the condensate to O($m$) in mass
perturbation theory in the $N=1$ case.\cite{Adam2}  Although the formula
(\ref{condensate1}) incorporates some of the effects of a fermion mass $m$
through  $\Bbar_a$ and $f(\vphi)$, it does not incorporate higher
order radiative corrections which become important for a large $m \gg \mu$.
We shall see that our formula gives a fairly good agreement with the 
lattice result for $m <  \mu$ in the $N=1$ case.  Adam's mass perturbation
theory  fails for $m > .5 \mu$.\cite{Harada2}

There is ambiguity in the definition of the composite
operator $\psibar \psi(x)$.  It diverges in perturbation theory.
It has to be normalized such that $\la \psibar \psi \ra = 0$
in a free theory ($e=0$) in the infinite volume limit 
($L\go \infty$).  In other words
\beeq
\la \psibar_a\psi_a\ra_\theta = 
\la \psibar_a\psi_a\ra_\theta' - \la \psibar_a\psi_a\ra_\free' ~.
\label{condensate2}
\eneq

The values of $\la \psibar_a\psi_a\ra_\theta'$ and $\la
\psibar_a\psi_a\ra_\free'$ depend on the regularization method
employed, but the difference does not.  In our regularization scheme we
shall find $\la \psibar_a\psi_a\ra_\free' 
= -  e^{2\gamma} m_a/\pi$. (See (\ref{zeroCondensate}).) 

The fermion mass term $H_\mass =\int dx \, \sum_a 
\big\{ m_a e^{i\delta_a}  M_{aa}  + ~({\rm h.c.}) ~\big\}$ 
has many effects.  In addition to giving
a ``potential'' in the zero mode sector as discussed in the previous section,
it also gives mass terms ($\propto \chi^2$) and other interactions. It
follows from (\ref{massOperator2}) that
\beeq
H_\mass \Rightarrow
 -{1\over L}  \int dx \sum_a
\bigg\{ m_a \Bbar_a \la e^{i(\vphi_a-2\pi p_W)/N} \ra_f 
  \prod_{\alpha=1}^N N_{\mu_\alpha} [ e^{iU_{\alpha a} \sqrt{4\pi}
\chi_\alpha}] +   ~({\rm h.c.}) ~ \bigg\}  ~~. 
\label{BosonMass1}
\eneq

When fermion masses are small compared with the coupling constant, it is
legitimate to expand $H_\mass$ in power series of
$\chi_\alpha$.   We define
\beeq
R_a + i I_a  
= {8\pi\over L} \, m_a \Bbar_a \cdot 
\la e^{ i (\vphi_a -  2\pi p_W/N)} \ra_f  ~.
\label{defineRI}
\eneq
It follows that
\beeq
H_\mass 
\Rightarrow \int dx \bigg\{ + {1\over \sqrt{4\pi}} 
\sum_\alpha\sum_a  \chi_\alpha U_{\alpha a} I_a 
+ {1\over 2} \sum_{\alpha\beta} \sum_a \chi_\alpha U_{\alpha a} R_a 
U^t_{a\beta} \chi_\beta \bigg\} ~~. 
\eneq
Including the additional mass term coming from the Coulomb interaction, 
one finds 
\beqn
H^\chi_\mass &=& \int dx ~ \bigg\{ 
+ {1\over \sqrt{4\pi}} \,  \chi_\alpha  U_{\alpha a} I_a 
+ {1\over 2} \, \chi_\alpha K_{\alpha\beta}\chi_\beta \bigg\}  \crn
\noalign{\kern 0pt}
K_{\alpha\beta} &=& {\mu^2\over N} \, \sum_{a,b} U_{\alpha a} U_{\beta b}
  + \sum_a U_{\alpha a} U_{\beta a} R_a  ~~~.
\label{BosonMass2}
\eeqn

$U_{\alpha a}$'s are determined such that $K_{\alpha\beta} = \mu^2_\alpha \,
\delta_{\alpha\beta}$.   In other words, we diagonalize
\beqn
{\cal K} &=& {\mu^2\over N} \pmatrix{ 1 &\cdots & 1 \cr
                                     \vdots & \ddots & \vdots\cr
                                     1 & \cdots & 1 \cr} 
  + \pmatrix{ R_1 &&\cr
              &\ddots&\cr
              &&R_N\cr}  ~~~, \crn
\noalign{\kern 6pt}
K &=&  U \, {\cal K} \, U^t 
  = \pmatrix{ \mu^2_1 &&\cr
              &\ddots&\cr
              &&\mu^2_N\cr} ~~.  
\label{BosonMass3}
\eeqn
The set of equations, (\ref{massOperator2}), (\ref{Schro4}), 
(\ref{defineRI}),  and (\ref{BosonMass3}) needs to
be solved simultaneously.  This is a Hartree-Fock approximation
applied to the zero mode   and oscillatory mode sectors.   We call it
the generalized Hartree-Fock approximation

In terms of $R_a$ the chiral condensates are, in case $\delta_a=0$, 
$m_a \la \psibar_a\psi_a\ra_\theta' 
= -  R_a/ 4\pi$.  
Eqs. (\ref{BosonMass3}) and (\ref{condensate2}) relate boson masses 
to chiral condensates.  It is a part of the PCAC (Partially Conserved Axial
Currents) relations.

As fermion masses become larger, nonlinear terms ($\sim \chi^n$) in $H_\mass$
become relevant.  The boson masses are not simply given by 
(\ref{BosonMass3}). Improvement is necessary.

\sxn{Massless fermions ($\alpha_a=\alpha$)}

When all fermions are massless and satisfy the same boundary conditions
$\alpha_a=\alpha$, the model is exactly solvable.  In this case
Eq. (\ref{Schro4}) reduces to
\beqn
\Bigg\{
- \bigg( {N\over 2\pi} \bigg)^2 {\dd^2\over \dd p_W^2} 
+ {(\mu L)^2\over 4}  \, p_W^2 
- (N-1) \Lap_\vphi \Bigg\} f(p_W, \vphi) = \ep
f(p_W,\vphi) &&\crn
\Lap_\vphi = \sum_{a=1}^{N-1} 
 {\dd^2 \over \dd \vphi_a^2} 
- {2\over N-1} \sum_{a<b}^{N-1} 
 {\dd^2\over \dd \vphi_a \dd \vphi_b} ~. \hskip 2cm &&
\label{Schro8}
\eeqn
The ground state, or the vacuum, wave function is independent of
$\vphi_a$.  It is given by
\beeq
f(p_W, \vphi) = {\rm const} \cdot e^{-\pi\mu L p_W^2/ 2N} ~~.
\label{masslessWF}
\eneq

The boson mass spectrum is given by $\mu_1\equiv \mu= \sqrt{N} e/\pi$ and
$\mu_2=\cdots=\mu_N=0$.  The orthogonal matrix $U$ in (\ref{chi-fields})
has $U_{1a} = 1/\sqrt{N}$ and $\Bbar_a$ in (\ref{massOperator2}) is
\beeq
\Bbar_a = \Bbar = B(\mu L)^{1/N} ~~.
\label{masslessBbar}
\eneq
For $N \ge 2$, $\la \cos(\vphi_a - 2\pi p_W /N) \ra_f=0$ in
(\ref{condensate1}) as $f(p_W, \vphi)$ is independent of  $\vphi_a$.
The chiral condensate $\la \psibar_a \psi_a \ra_\theta$ vanishes 
for $N \ge 2$, reflecting Coleman's theorem which states that
in two dimensions a continuous symmetry cannot be spontaneously
broken.\cite{Coleman3}  The non-vanishing $\la \psibar_a \psi_a \ra_\theta$
breaks the $SU(N)$ chiral symmetry. For $N=1$ the $U(1)$ chiral symmetry is
broken by anomaly. Recalling $\vphi_1= \theta$ in (\ref{potentail1}) for
$N=1$, we have
\beeq
\la \psibar_a \psi_a \ra_\theta = \cases{
- \myfrac{2}{L} \, B(\mu L) \, e^{-\pi/\mu L} \, \cos \theta &for $N=1$\cr
0 &for $N\ge 2$.\cr}
\label{masslessCondensate1}
\eneq
The condensate for $N=1$ is plotted in fig.\ 1.

\begin{figure}[bt]
\vskip 0.3cm
\hskip 3cm
\epsfxsize= 9.5cm
\epsffile[39 290 486 601]{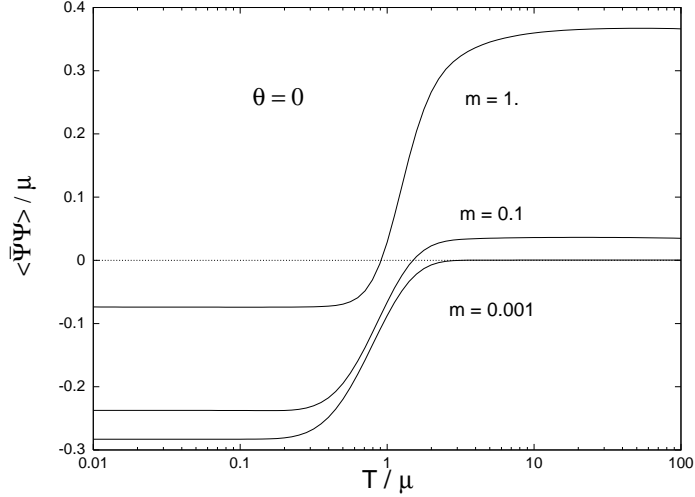}
\vskip -0cm 
\caption{\small Temperature dependence of the chiral condensate in the $N=1$ 
model at $\theta=0$.  There appears crossover transition around $T/\mu=1$.}
\label{fig.1}
\end{figure}

The $SU(N)$ invariant condensate is non-vanishing, however.  It follows from
(\ref{state1}) that
\beeq
\la p_W', n',\vphi' | \exp \Big( \pm i \sum_{a=1}^N q_a \Big)
| p_W, n, \vphi \ra 
= \delta_{n', n\pm 1} \delta (p_W'-p_W) 
   \prod_{a=1}^{N-1} \delta_{2\pi}(\vphi_a'-\vphi_a) ~.
\label{qMatrix5}
\eneq
Straightforward calculations in the Schr\"odinger picture yield
\beqn
\la \psibar_{NL} \cdots \psibar_{1L} \psi_{1R} \cdots \psi_{NR} \ra_\theta
&=&  \Big({-1\over L}\Big)^N \la e^{-i\sum_a q_a }  
     N_0[ e^{i\sqrt{4\pi N} \chi_1} ] \ra_\theta  \cr
\noalign{\kern 10pt}
&=& \bigg[{-B(\mu L)\over L}\bigg]^N  e^{-i\theta} e^{-N\pi/\mu L}
~~. 
\label{masslessCondensate2}
\eeqn
In the infinite volume limit it approaches
$ e^{-i\theta} \big( - \mu e^\gamma / 4\pi \big)^N$.

\vskip .5cm

\sxn{Truncation}

There are two potential terms in the equation for $f(p_W, \vphi)$, 
Eq.\ (\ref{Schro4}).  If fermion masses are small compared to 
$\mu$, one of them, $(\mu L)^2 p_W^2/4$, dominates over the 
other term in $V(p_W,\vphi)$.  The condition is $m_a \Bbar_a \ll \mu^2 L$.
For $m_a \ll \mu$, it is satisfied when $\mu L\gg m/\mu$ 
($T/\mu \ll \mu/m$).

In such situation $P_W$ acts as a fast variable, whereas $\vphi_a$'s 
act as slow variables.  The wave function can be approximated by
$f(p_W,\vphi) = \sum_{s=0}^\infty u_s(p_W) f_s(\vphi) \sim u(p_W) f(\vphi)$
where
$u(p_W) = u_0(p_W)$ is 
\beqn
&&\Bigg\{
- \bigg( {N\over 2\pi} \bigg)^2 {d^2\over d p_W^2} 
+ {(\mu L)^2\over 4}  \, p_W^2   \Bigg\} ~ u(p_W) 
=  {N\mu L\over 4\pi}  ~ u(p_W) \crn
&&\hskip 2cm u(p_W) 
= \Big( {\mu L\over N} \Big)^{1/4}\, e^{-\pi\mu L p_W^2/2N}
\label{Schro5}
\eeqn
Then cosine term in $V_N(p_W, \vphi)$ is approximated by
\beqn
 \cos \Big(\vphi_a - {2\pi p_W\over N} \Big) &\go& 
\int dp_W \, \cos \Big(\vphi_a - {2\pi p_W\over N} \Big) \, u(p_W)^2 \crn
&=& e^{-\pi/N\mu L} \, \cos \vphi_a ~~~.
\label{factorize1}
\eeqn

We remark that the truncated equation has symmetry
$p_W \go - p_W$,  though the original equation  does not in general. 
The truncation may not be good for physical quantities sensitive to
this symmetry.

The function $f(\vphi)$ satisfies
\beqn
&&\bigg\{ - (N-1) \Lap_\vphi + V_N(\vphi) \bigg\} f(\vphi) 
    = \ep_0 f(\vphi) \crn 
&&V_N(\vphi) = - {NL\over \pi}\,  e^{-\pi/N\mu L}\,
    \sum_{a=1}^N m_a  \Bbar_a \,  \cos \vphi_a  
\label{Schro6}
\eeqn
This equation has been extensively studied in the $N$=2 and $N$=3 cases in
Refs \cite{HHI,HHI2}.

Let us denote $\lla F(\vphi) \rra_f = \int [d\vphi] ~  F(\vphi) ~
|f(\vphi)|^2$.
Then (\ref{qMatrix3}) and (\ref{defineRI}) become
\beqn
&&\la e^{\pm i q_a} \ra_\theta = e^{\mp i \delta_a} \, e^{-\pi/N\mu L} \,
\lla e^{\pm i \vphi_a} \rra_f \crn
&&R_a +i I_a 
= {8\pi\over L}  \,m_a \Bbar_a \, e^{-\pi/N\mu L}
 \lla  e^{i\vphi_a} \rra_f   ~.
\label{defineRI2}
\eeqn
All formulas in Section 4 remain intact with these substitutions.
In particular, the chiral condensate satisfies
$\la \psibar_a\psi_a \ra_\theta' = - R_a/4\pi m_a$.

\vskip .5cm

\sxn{$N=1$  massive case}

\noindent
(1) Generalized Hartree-Fock approximation

With one fermion there is no $\vphi$ degree of freedom.  
We write $m_1=m, \delta_1=0, \alpha_1=\alpha$.  
Recall that $\mu= e/\sqrt{\pi}$ and $\vphi_1=\theta_\eff $. The vacuum
wave function is determined  by
\beqn
|\Psi_\vac(\theta) \ra = {1\over \sqrt{2\pi}} \sum_n \int dp_W 
~ |p_W, n \ra  ~e^{-in\theta + 2\pi i (n+\alpha) p_W}
    ~  f(p_W) \hskip 2cm &&\crn 
\Bigg\{  -  {1\over (2\pi)^2}  {d^2\over d p_W^2} 
+{(\mu L)^2\over 4}  \, p_W^2
-  { m L B(\mu_1 L)\over \pi}  \cos (2\pi p_W - \theta_\eff )
  \Bigg\}  f(p_W) = \ep f(p_W) ~.&&
\label{N=1vacuum1}
\eeqn
When $m \ll \mu$,  the boson mass $\mu_1$ must satisfy
\beeq
\mu_1^2 = \mu^2 + {8\pi m B(\mu_1L)\over L} 
 ~  \la \cos (2 \pi p_W - \theta) \ra_f~~.
\label{N=1BosonMass}
\eneq
As $m$ becomes larger, the formula (\ref{N=1BosonMass}) needs to be 
improved to incorporate nonlinear effects in the fermion mass $m$.
In particular,  $\mu_1 = 2 m + {\rm O}(e^2/m)$ for $m\gg \mu$, as the 
boson is interpreted as a fermion-antifermion bound-state.

In determining a physical chiral condensate, one needs to subtract
a condensate in free theory as discussed in Section 4.  The
``free'' limit corresponds to the limit $m\gg \mu$.  At the moment we do
not have 
reliable formulas which relate $\mu_1$, $m$, and $\la\psibar\psi\ra_\theta'$
for $m \gg \mu$.  The best we can do is to extrapolate (\ref{condensate1})
and (\ref{N=1BosonMass}) for large $m$ to determine the subtraction
term within our approximation.

It follows from (\ref{N=1vacuum1}) that
 in the week coupling  $e/m \ll 1$ and infinite volume
$L\go\infty$ limits $\la \cos(2\pi p_W - \theta) \ra_f = 1$.   
If the formula  (\ref{N=1BosonMass}) is employed, one obtains
$\mu_1 = 2 e^\gamma m$.  Combined with (\ref{condensate1}), it gives
\beeq
\la \psibar \psi \ra_\free' = - {e^{2\gamma}\over \pi} \, m ~.
\label{zeroCondensate}
\eneq
The chiral condensate is therefore given by
\beeq
\la \psibar \psi\ra_\theta = - {2 B(\mu_1L)\over L} 
 ~  \la \cos (2 \pi p_W - \theta ) \ra_f  + {e^{2\gamma}\over \pi} m 
\label{N=1condensate1}
\eneq
which we  expect to be a good approximation for $m < \mu$.
We stress that within our approximation the subtraction term is given 
by (\ref{zeroCondensate}).  In an exact treatment the boson mass
in the weak coupling limit should be given by $\mu_1=2m$. To achieve
this, one has to improve both (\ref{condensate1}) and
(\ref{N=1BosonMass}) consistently.  

In the massless case $(m=0)$, 
$\la \psibar \psi\ra_\theta^{m=0}
   = - 2 L^{-1} B(\mu L)  \, e^{-\pi/\mu L} \, \cos\theta$. 
In the infinite volume limit  or zero temperature limit, 
it approaches $ - (\mu e^\gamma /2\pi)  \, \cos\theta$.

It may be of interest to apply a perturbation theory \cite{massive} to
(\ref{N=1vacuum1}) when $m\ll \mu$. Write (\ref{N=1vacuum1}) in the form
\beqn
&&(H_0 + V) \, | \Psi \ra = E \, | \Psi \ra \crn
&&H_0 = 2\pi\mu L (a^\dagger a + \onehalf) \crn
&&V = - \kappa \cos(2\pi p_W -\theta) ~~,~~
\kappa = 4\pi mL B(\mu_1 L) 
\label{N=1vacuum2}
\eeqn
where the annihilation and creation operators are 
\beeq
 \pmatrix{a \cr a^\dagger \cr} =
{1\over \sqrt{2}} \Big( \pm {1\over \sqrt{\pi\mu L}} {d\over dp_W}
  + \sqrt{\pi\mu L} \, p_W \Big)  ~~.
\label{N=1vacuum3}
\eneq
In terms of number eigenstates $|n\ra$,
$|\Psi\ra$ is found to be, to O($m/\mu$),
\beeq
|\Psi\ra = |0\ra - {1\over 2\pi\mu L} \sum_{n=1}^\infty
    {1\over n} \, |n\ra \la n | V|0\ra ~~.
\label{N=1WF1}
\eneq

Since $\cos(2\pi p_W - \theta) = \onehalf e^{-\pi/\mu L} 
\Big( e^{-i\theta} e^{i\alpha a^\dagger} e^{i\alpha a} + {\rm h.c.} \Big)$
where $\alpha = (2\pi / \mu L)^{1/2}$, 
it follows that
$\la \cos(2\pi p_W-\theta) \ra_f^{(0)} =e^{-\pi/\mu L} \, \cos\theta$ and
\beqn
&&\la \cos(2\pi p_W-\theta) \ra_f^{(1)} = 
{\kappa\over \pi \mu L} \sum_{n=1}^\infty
{1\over n} \Big| \la 0 | \cos(2\pi p_W - \theta) |n \ra \Big|^2 \crn
&&= {\kappa\over 2\pi\mu L} \, e^{-2\pi/\mu L}
\sum_{n=1}^\infty {1+ (-1)^n \cos 2\theta\over n \cdot n!} ~
  \Big( {2\pi\over \mu L} \Big)^n \crn
&&= {\kappa\over 2\pi\mu L} \, e^{-2\pi/\mu L}
\bigg\{ \int_0^{2\pi/\mu L} dz \, {e^z - 1\over z}
+ \cos 2\theta \int_0^{2\pi/\mu L} dz \, {e^{-z} - 1\over z} \bigg\}\crn
&&= e^\gamma \, {m\over \mu} \, (1- \cos 2\theta)
  \hskip 1cm {\rm for} ~ \mu L \gg 1  ~.
\label{N=1identity3}
\eeqn

The boson mass $\mu_1 = \mu + \delta\mu$ is obtained from
(\ref{N=1BosonMass}):
\beqn
\delta\mu &=&  {4\pi\over \mu L} \, m B(\mu L) e^{-\pi/\mu L}
\cos\theta  + {\rm O} (m^2/\mu^2) \crn
&\sim& me^\gamma \cos\theta \hskip 1cm {\rm for~} \mu L \gg 1~~. 
\label{N=1BosonMass2}
\eeqn
The chiral condensate is given by
\beqn
\la \psibar\psi \ra_\theta' &=&
-{2\over L}\Big\{ B(\mu L) 
    + B'(\mu L) \delta\mu L \Big\} e^{-\pi/\mu L} \cos\theta   
 - {2\over L} B(\mu L) \la \cos(2\pi p_W-\theta) \ra_f^{(1)} \crn
&=& -{e^\gamma\over 2\pi} \, \mu \cos\theta + 
{e^{2\gamma}\over 4\pi} (-3 + \cos 2\theta) m
\hskip 1cm {\rm for ~} \mu L \gg 1 ~.
\label{N=1condensate4}
\eeqn
With (\ref{zeroCondensate}) 
\beeq
\la \psibar\psi \ra_\theta \sim
-{e^\gamma\over 2\pi} \, \mu \cos\theta + 
{e^{2\gamma}\over 4\pi} (1 + \cos 2\theta) m ~.
\label{N=1condensate5}
\eneq
This result differs from the result in the mass perturbation theory for 
the reason explained above and should not be taken seriously.
Our formalism, however, allows us to estimate 
$\la \psibar\psi\ra_\free$, and therefore the physical condensate
$\la \psibar\psi\ra_\theta$.  The
mass perturbation theory is valid only for small $m/\mu \ll 1$, whereas
our formalism allows the numerical evaluation of various
physical quantities even for $m/\mu
\sim 1$. We shall see below a good agreement between ours and the
lattice gauge theory in the range $m/\mu < 1$.  See the subsection (3)
below.

\bigskip

\noindent (2)  Mass perturbation theory

Corrections can be evaluated in a power series in $m/e$.  This analysis
was carried out at zero temperature  by
Adam.\cite{Adam2} We present the analysis at finite temperature.

We illustrate the computation for the chiral condensate. Recall
$\psibar\psi = - B(\mu L) L^{-1} ( e^{iq} K_+ + e^{-iq} K_-)$
where $K_\pm  =
e^{\pm i\sqrt{4\pi} \phi^{(-)} }e^{\pm i\sqrt{4\pi} \phi^{(+)}}$. 
We have suppressed 
irrelevant factors.   In the invariant perturbation theory
\beeq
\la \psibar\psi \ra_c^{(1)} = - i m \bigg\{ {B(\mu L)\over L}
\bigg\}^2
\int d^2 x \, \sum_{a,b=\pm}
\la T[e^{iaq(t)} e^{ibq(0)}] T[K_a(x) K_b(0) ]\ra_c ~~.
\label{condensate3}
\eneq
Making use of (\ref{formula-2}) and the identity
$\la e^{i\alpha \phi^{(+)}(x)} e^{i \beta\phi^{(-)}(y)} \ra
= e^{-\alpha\beta [\phi^{(+)}(x), \phi^{(-)}(y)]}$, one finds
\beqn
\la \psibar\psi \ra_c^{(1)} &=&
-{2im \over L^2} B(\mu L)^2 e^{-2\pi/\mu L}
\int d^2 x \, \Big\{ (e^{4\pi i G(x)} -1)
   + \cos 2\theta \, (e^{-4\pi iG(x)} -1) \Big\} \cr
\noalign{\kern 10pt}
G(x)  &=& {1\over 2\pi L} \sum_n \int_{-\infty}^\infty d\omega \, 
{e^{-i\omega t + i p_n x} \over \omega^2 - p_n^2 -\mu^2 + i\ep}  ~~.
\label{condensate4}
\eeqn  
The disconnected component has been subtracted.  Deforming the 
$t$-integral to the imaginary axis, one finds
\beqn
&&\la \psibar\psi \ra_c^{(1)} =
-{m \over \pi^2} B(\mu L)^2 e^{-(2\pi/\mu L)} \cr
&&\hskip 2cm \times
\int_0^\infty d\tau \int_0^{2\pi}  dx \,
\Big\{ ( e^{+E(\tau,x; \mu L/2\pi)} -1)
  + \cos 2\theta \, (e^{- E(\tau,x; \mu L/2\pi)} -1) \Big\} \cr
\noalign{\kern 10pt}
&&\hskip 1.5cm \equiv [F_+ (\mu L) 
+ F_-(\mu L) \cos 2\theta ] m \cr
\noalign{\kern 10pt}
&&E(\tau,x; z) = {1\over z} \, e^{- z \tau}
+ 2 \sum_{n=1}^\infty {1\over v_n}
e^{-v_n\tau}  \, \cos nx \next v_n = (n^2 + z^2)^{1/2} ~~~.
\label{condensate5}
\eeqn

The coefficient $F_-(\mu L)$ is finite, whereas 
$F_+(\mu L)$ diverges logarithmically near $\tau=x=0$.  The divergence is
due to the O($m$) correction in the  free ($e=0$) theory, which must be
subtracted to define the physical chiral condensate as explained in
Section 4.  Hence in the mass perturbation theory
\beqn
\la \psibar \psi \ra_\theta &=&
-{e^\gamma\over 2\pi} \mu \cos\theta
+ m \Big[   F_+(\mu L) - F_+^{\rm free} + F_-(\mu L) \cos 2\theta \Big] 
+ \cdots ~. 
\label{condensate6}
\eeqn
$F_+^{\rm free}$ is obtained in a similar manner. One starts with a
massless free fermion theory.  In the bosonization method
the vacuum satisfies $p | \vac \ra =0$.  Since
$\la  e^{\pm iq}\ra =0$,  the condensate vanishes; $\la \psibar \psi \ra
=0$.  The fermion mass is treated as  a perturbation.  To O($m$) one gets an
expression which is the same as (\ref{condensate3}) except that $B(\mu L)$
is replaced by 1 and 
$\phi(x)$ represents a massless field.  Making use of (\ref{formula-f2}),
one finds that 
\beeq
\la \psibar\psi \ra^{(1)}_{\rm free} 
= - {2im\over L^2} \int d^2 x \, e^{-2\pi i |t|/L}
 e^{4\pi i G_{\rm free}(x) }
\label{free-condensate1}
\eneq
where $G_{\rm free}(x)=-i\la T[\phi(x) \phi(0)]\ra$ is a massless
propagator excluding the contribution from the zero mode.
Employing (\ref{formula2}) and deforming the integration path, one finds
\beeq
\la \psibar\psi \ra^{(1)}_{\rm free} 
= - {m\over \pi^2} \int_0^\infty d\tau \int_0^{2\pi} dx \, 
{ e^{-\tau} \over (1- e^{-\tau-ix})(1- e^{-\tau +ix}) } 
\equiv m \, F_+^{\rm free}   ~~.
\label{free-condensate2}
\eneq
Notice that $F_+^{\rm free}$ is independent of $L$.  Comparing 
(\ref{free-condensate2}) with (\ref{condensate5}), we observe that
$F_+^{\rm free} =  F_+(\mu L) \big|_{\mu L \go 0}$.

$F_-(\mu L)$ can be easily evaluated by numerical integration.
At $\mu L/2\pi=10 \sim 50$, $F_-= 0.357$.  This is consistent with the
number, 0.3581, obtained by Adam in the $L\go\infty$ limit.  
The divergence  in $F_+(\mu L)$ and $F_+^{\rm free}$  makes the evaluation
of the difference very difficult. We comment that  Adam's subtraction
procedure to get finite 
$F_+^{\rm free}$ is inconsistent.  Indeed, his massless propagator
differs from that obtained by taking the $\mu\go 0$ limit of
the massive propagator in ref.\ \cite{Adam2}.
Further, his numerical estimate,  $-0.39126$, for 
$F_+(\infty) - F_+^{\rm free}$  disagrees with the lattice
result.  At $\theta=0$ Adam's estimate gives
$F_+(\infty) - F_+^{\rm free} + F_-(\infty) = - 0.0345 <0$ in Eq.\
(\ref{condensate6}), which contradicts with the recent result from the lattice
gauge theory.  (See fig.\ 2.) There is a disagreement in the sign.

\bigskip \noindent
(3)  Numerical evaluation in the generalized Hartree-Fock approximation

In this subsection we present various results obtained by
numerical evaluation.  The algorithm is simple.  With given $\mu L$,
$m/\mu$, and $\theta_\eff$, we start with an initial $\mu_1/\mu$.  Then
Eq. (\ref{N=1vacuum1}) is solved numerically to find $f(p_W)$.
With this $f(p_W)$, a new $\mu_1/\mu$ is determined by (\ref{N=1BosonMass}).
We have a mapping
\beeq
\mu_1 \go f(p_W) \go \mu_1 ~~.
\nonumber
\eneq
We repeat this process until the output $\mu_1$ coincides with the 
input $\mu_1$ within required accuracy.  With $\mu_1$ being fixed, the 
condensate is evaluated by (\ref{N=1condensate1}).

In fig.\ 2 the 
$m$ dependence of the condensate is plotted at $\theta=0$ with
various values of $T/\mu$.  The lattice result from ref.\ \cite{deForcrand}
is also plotted for comparison.  Our result agrees well with 
Tomachi and Fujita's evaluation by the Bogoliubov
transformation.\cite{Tomachi} The agreement with the lattice result is
modest. Note that the subtraction of condensates in free
theory in each regularization scheme is crucial.

\begin{figure}[bt]
\vskip 0.3cm
\hskip 3cm
\epsfxsize= 9.5cm
\epsffile[39 190 484 503]{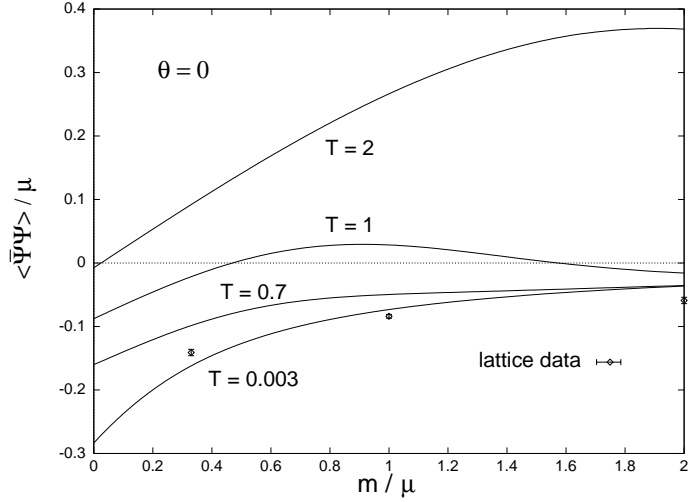}
\vskip -0cm 
\caption{\small The mass dependence of the chiral condensate in the $N=1$
model at $\theta=0$ with various values of temperature $T$.  In the figure
$T$ is in a unit of $\mu$.  The lattice data is by de Forcrand et
al.\cite{deForcrand}  The additional lattice data point at $m/\mu=2$ was
provided by the authors of ref.\ \cite{deForcrand}. The condensate for
$T/\mu < 0.1$ is essentially the same as  that at $T=0$.  The curve for
$T/\mu=0.003$ is consistent with the result by Tomachi and
Fujita.\cite{Tomachi}}
\label{fig.2}
\vskip 0.3cm
\end{figure}

There appears  a singularity in the $m$ dependence of the condensate when
$\theta$ is close to $\pi$.  In fig.\ 3 the condensates are plotted
with various values of $\theta$.   The discontinuity appears at 
$m/\mu = .44$ for $\theta=\pi$, and at $m/\mu = .40$ for
$\theta=.95 \pi$.

\begin{figure}[bt]
\vskip 0.3cm
\hskip 3cm
\epsfxsize= 9.5cm
\epsffile[29 190 474 503]{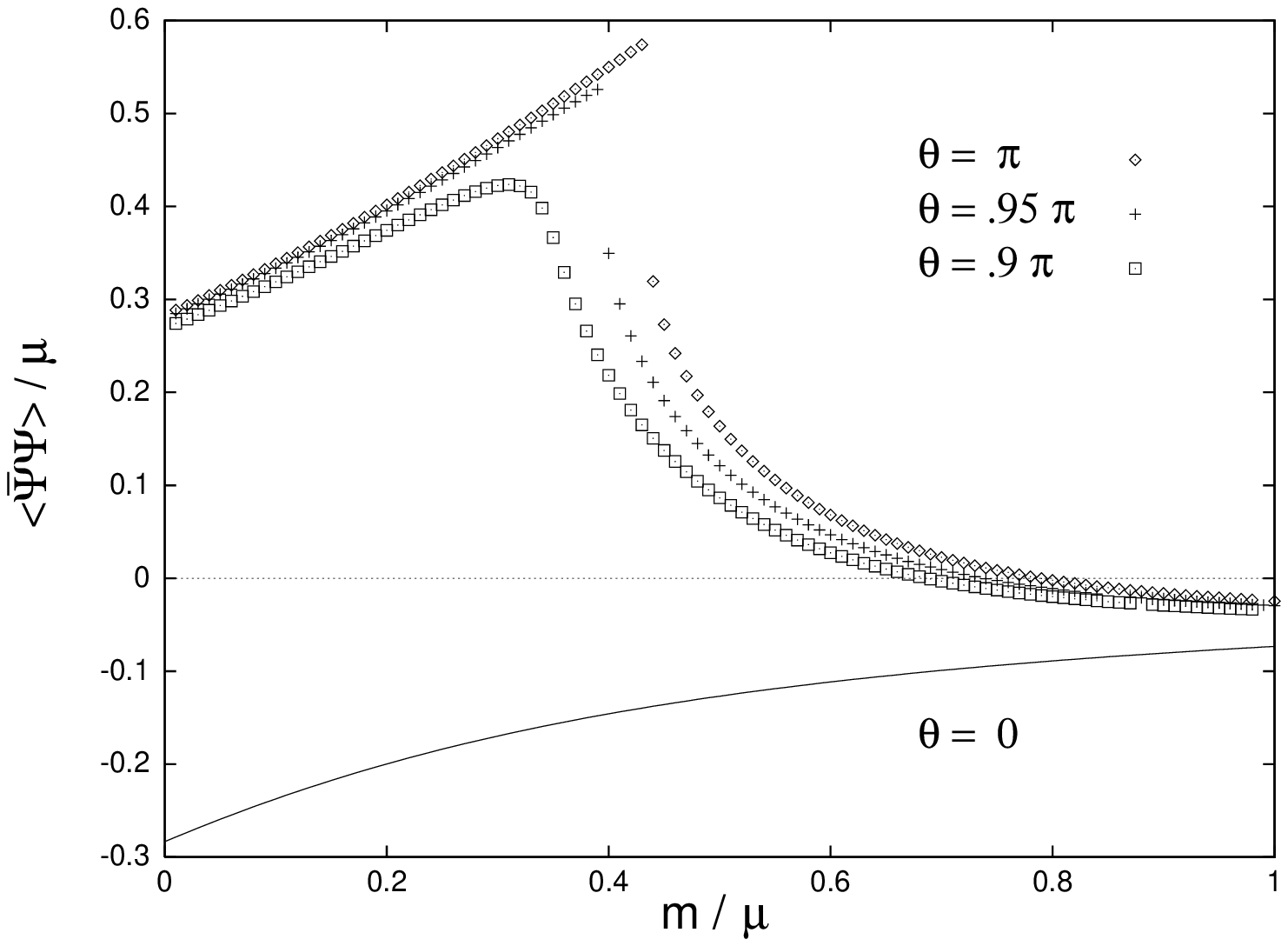}
\vskip -0cm 
\caption{\small The mass dependence of the chiral condensate in the $N=1$
model near $\theta = \pi$ at $T/\mu=0.003$. There appears a discontinuity
above  $\theta \sim .95 \, \pi$.  The condensate at $\theta=0$ is also 
displayed for comparison.}
\label{fig.3}
\end{figure}

The discontinuity persists so long as the temperature is lower than 
$T_c \sim .12 \mu$.
The condensate at various values of $T$ is depicted in fig.\ 4.

\begin{figure}[bt]
\vskip 0.3cm
\hskip 3cm
\epsfxsize= 9.5cm
\epsffile[45 190 474 503]{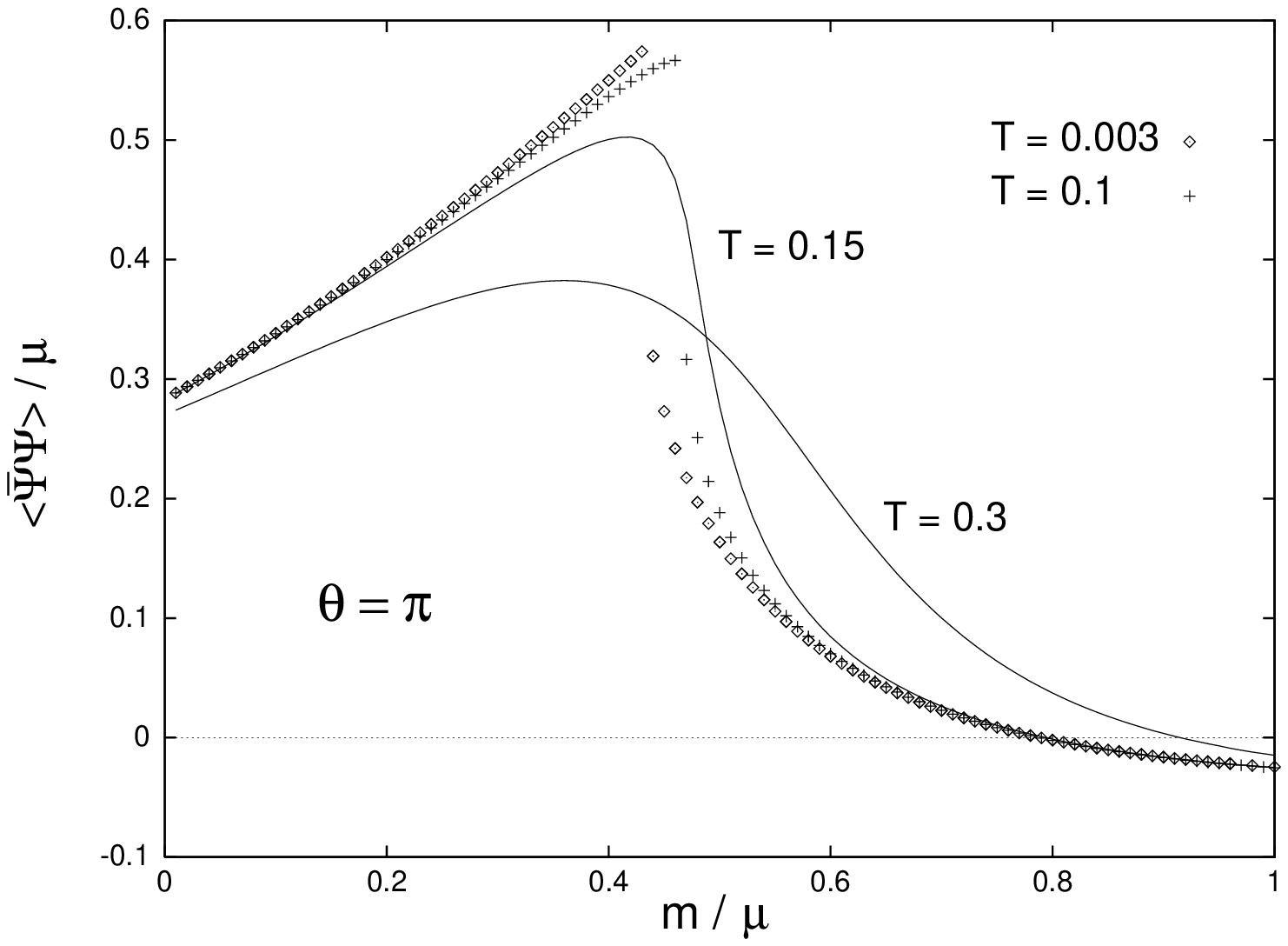}
\vskip -0cm 
\caption{\small The mass dependence of the chiral condensate in the $N=1$
model at $\theta = \pi$ at various values of $T/\mu$. For $T/\mu < 0.12$
the discontinuity remains, whereas the transition becomes smooth
for $T/\mu > 0.12$.}
\label{fig.4}
\end{figure}

The origin of the discontinuity is understood as follows.\cite{massive}
For $\mu L\gg 1$ Eq. (\ref{N=1vacuum1}) becomes
\beqn
\Bigg\{  -  {1\over (2\pi)^2}  {d^2\over d p_W^2} 
+{(\mu L)^2\over 4}  \, p_W^2
-  { m\mu_1 L^2e^\gamma\over 4\pi^2}  \cos (2\pi p_W - \theta_\eff )
  \Bigg\}  f(p_W) = \ep f(p_W) ~~~.
\label{N=1vacuum4}
\eeqn
The potential term dominates over the kinetic energy term.  Suppose that
$\theta_\eff=\pi$.  Then the wave function is sharply localized around
the absolute minimum of $\hat V(p_W) = (\pi\mu p_W)^2 + m\mu_1 e^\gamma
\cos(2\pi p_W)$.  There is always  a  solution for which
$\mu^2 > 2 e^\gamma m \mu_1$.   In this case $\hat V(p_W)$ is minimized at 
$p_W=0$ and $\mu_1=\sqrt{\mu^2 +m^2 e^{2\gamma}} - me^\gamma$.  There is 
another solution for $m/\mu > 0.435$, or $\mu^2 < 2 e^\gamma m \mu_1$.  
$\hat V(p_W)$ is minimized at $p_W=\bar p_W$ where 
$2\pi\bar p_W = (2m\mu_1 e^\gamma/\mu^2) \sin 2\pi \bar p_W$.  The boson
mass is given by $\mu_1 = \sqrt{\mu^2 + a^2} - a$ where $a=me^\gamma \cos 2\pi
\bar p_W$.  The second solution has a lower energy density and corresponds
to the vacuum.  Hence at $T=L^{-1}=0$, there appears a discontinuity at 
$m_c/\mu=0.435$.

At finite temperature the critical value  $m_c/\mu$ is determined
numerically.  $m_c/\mu = 0.437$ at $T/\mu=0.03$.  The discontinuity
disappears for $T/\mu > 0.12$.  As noted  in ref. \cite{massive} 
there are two possible
scenarios.  In the full theory the discontinuity may persist, but with 
a universal $m_c/\mu$ independent of $T/\mu$.  Or the discontinuity
may be smoothed by higher order corrections.  At the moment we do not know
for sure which picture is right.

\vskip 1cm

\sxn{Degenerate fermions}

When all fermion have degenerate masses ($m_a=m$) and obey the 
same boundary conditions $\alpha_a=\alpha$, the Laplacian 
$\Lap_\vphi$ and potential $V_N(p_W, \vphi)$ in the eigenvalue
equation (\ref{Schro4}) become
\beqn
\Lap_\vphi &=& \sum_{a=1}^{N-1} {\dd^2\over \dd \vphi_a^2}
- {2\over N-1} \sum_{a<b}^{N-1} {\dd^2\over \dd \vphi_a\dd \vphi_b} \crn
V(p_W, \vphi) &=& + {(\mu L)^2\over 4}  \, p_W^2
- {NmL\Bbar\over \pi} 
\sum_{a=1}^N \cos\Big( \vphi_a - {2\pi p_W\over N} \Big) \crn
\Bbar &=& B(\mu_1 L)^{1/N} B(\mu_2 L)^{1 - (1/N)} ~~.
\label{sym-potential1}
\eeqn
Boson masses $\mu_1$ and $\mu_2= \cdots = \mu_N$ are determined by
(\ref{defineRI}) - (\ref{BosonMass3});
\beqn
&&\mu_1^2 = \mu^2 + R  ~~,~~ \mu_2^2 = R \crn
&&R = {8\pi m \Bbar\over L} \,
\la \cos \Big(\vphi_a - \myfrac{2\pi p_W}{N}\Big) \ra_f ~~.
\label{sym-boson1}
\eeqn

When $\mu L, \mu_1 L , \mu_2 L \gg 1$, the potential is approximated by
\beeq
V_\infty = {L^2\over 4\pi^2} \, \Bigg\{ (\pi \mu p_W)^2 
- N e^\gamma \, m \mu_1^{1/N} \mu_2^{(N-1)/N} 
\sum_{a=1}^N \cos\Big( \vphi_a - {2\pi p_W\over N} \Big) \Bigg\}~~.
\label{sym-potential2}
\eneq

If fermion masses are small $m/\mu \ll 1$, the first term
in the potential dominates over the second.  The $p_W$ dependence of
the wave function is the same as in the massless case, and one can 
write $f(p_W, \vphi)= e^{-\pi \mu LP_W^2/2N} f(\vphi)$.
With the aid of the truncation formula 
(\ref{factorize1}) the equation is reduced to
\beqn
&&\Big\{ - \Lap_\vphi + \kappa  F_N(\vphi)
\Big\} \, f(\vphi) = \ep \, f(\vphi)  \crn
&&\kappa =  \myfrac{N}{\pi(N-1)} mL\Bbar   e^{-\pi/N\mu L} \next
F_N(\vphi) = - \sum_{a=1}^N \cos \vphi_a ~~.
\label{sym-Schro2}
\eeqn
The boson mass $\mu_2$ is given by
\beeq
\mu_2^2  = {8\pi m \Bbar\over L} \, e^{-\pi/N\mu L} \,
\lla \cos \vphi_a \rra_f 
= {8\pi^2 (N-1)\over NL^2} \, \kappa \, \lla \cos \vphi_a \rra_f~.
\label{sym-mass1}
\eneq

In the $L\go\infty$ or $T\go 0$ limit, the wave function has a 
sharp peak at the location of the minimum of $V_\infty(p_W,\vphi)$ in
(\ref{sym-potential2}) or $F_N(\vphi)$  in (\ref{sym-Schro2}).
We examine the location of the minimum of the potential.

\bigskip

\leftline{\bf (1) Potential}
\bigskip
\leftline{(a) $N=2$}
In the two flavor case $\vphi_1=\vphi, \vphi_2=\theta-\vphi$ and the
potential takes the form of
\beeq
V_\infty(p_W, \vphi) = {L^2\over 4\pi^2} \, \Bigg\{ (\pi \mu p_W)^2 
- 4 e^\gamma \, m \mu_1^{1/2} \mu_2^{1/2} 
\cos\Big( \pi p_W - {\theta\over 2} \Big) 
  \cos \Big( \vphi - {\theta\over 2} \Big) \Bigg\}
\label{Vpot2}
\eneq
or
\beeq
F_2(\vphi) = - 2 \cos{\theta \over 2} \cos \Big( \vphi - {\theta\over
2}\Big) ~.
\label{Fpot2}
\eneq
The form of the potential $V_\infty$ suggests that the anomalous
behavior analogous to that in the $N=1$ case may develop near
$\theta=\pi$ with $m={\rm O}(\mu)$.  Note that at $\theta=\pm\pi$
the truncated equation (\ref{sym-Schro2}) is not valid as the 
potential $F_2(\vphi)=0$.  The $p_W$ degree of freedom must be retained.

$F_2(\vphi)$ has a minimum at
\beeq
\vphi = {\bar\theta\over 2} \next
 \bar\theta = \theta - 2\pi \bigg[ {\theta+\pi\over 2\pi} \bigg]
\label{minimum1}
\eneq
where $[x]$ denotes a maximum integer not exceeding $x$ so that 
$|\bar\theta| \le \pi$.  Notice that
the location of the minimum discontinuously changes at $\theta=\pi$
($mod ~ 2\pi$).

\bigskip
\leftline{(b) $N=3$}
Let us examine the potential $F_3(\vphi)$. 
\beqn
F_3(\vphi) = - \cos(\vphi_1+\vphi_2-\theta) - \cos\vphi_1
-\cos\vphi_2  ~~.  
\eeqn
The location of the minimum can be easily found analytically.  
\ignore{ Since
${\dd F/\dd \vphi_1}\propto \sin\onehalf(2\vphi_1+\vphi_2-\theta)
\cos\onehalf(\vphi_2-\theta)$  etc., all stationary points of the
potential must satisfy 
(i) $2\vphi_1+\vphi_2-\theta=0$ or $\vphi_2-\theta=\pi$,  and 
(ii) $\vphi_1+2\vphi_2-\theta=0$ or $\vphi_1-\theta=\pi$ ($mod~2\pi$).}
There are six distinct stationary points:
$\vphi_1=\vphi_2=\onethird \theta, \onethird (\theta \pm 2\pi)$,
or $(\vphi_1,\vphi_2) =(\theta+\pi,\theta+\pi)$, $(\theta+\pi, -\theta)$,
$(-\theta, \theta+\pi)$.   The global minimum is located at
$\vphi_1=\vphi_2 ={1\over 3} \bar\theta$.
The periodicity in $\theta$ is $2\pi$.   
The location of the minimum jumps from 
$(\vphi_1,\vphi_2)=(+{1\over 3}\pi , +{1\over 3}\pi) $ to 
$(-{1\over 3}\pi , -{1\over 3}\pi) $ at $\theta=\pi ~~({\rm mod}~ 2\pi)$. 
The minimum is always located at $|\vphi_1|=|\vphi_2| \le {1\over 3} \pi$.

\bigskip 
\leftline{(c) General $N$}
First notice that
\beqn
&&F_N(\vphi_1, \cdots,\vphi_{N-1}; \theta)
= - \sum_{a=1}^{N-1} \cos \vphi_a 
  - \cos \Big( \theta- \sum_{a=1}^{N-1} \vphi_a \Big) \cr
&&= F_{N-1}(\vphi_1, \cdots,\vphi_{N-2}; \theta - \vphi_{N-1})
   - \cos \vphi_{N-1}  ~~. 
\label{FpotentialN}
\eeqn
For $N=3$ we know the minimum is located at $\vphi_1=\vphi_2={1\over
3}\bar\theta$. 

Consider $N=4$.  We denote the minimum of $F_4$ by 
$(a_1,a_2,a_3)$.  Fix  the value of $\vphi_3$ and consider
$F_4(\vphi)_{\vphi_3 {\rm fixed}}\equiv G_4(\vphi_1,\vphi_2)$.   
Denote the minimum of $G_4$ by $(b_1,b_2)$ where $b_j=b_j(\vphi_3)$.
It follows from  (\ref{FpotentialN}) that $(b_1,b_2)$ is the minimum of 
$F_3(\vphi_1,\vphi_2;\theta-\vphi_3)$.  The result in the $N=3$ case
implies  that  $b_1=b_2$.
Since the minimum of $H_4(\vphi_3) = G_4(b_1[\vphi_3],b_2[\vphi_3];\vphi_3)$,
which we denote by $c$,
is the minimum of  $F_4(\vphi)$,  we have $[a_1,a_2,a_3] = [b_1(c), b_2(c),
c]$.  In particular,  $a_1=a_2$ as $b_1=b_2$.
We repeat the argument with the value of $\vphi_1$ kept fixed, to obtain
$a_2=a_3$.  Hence $a_1=a_2=a_3$, i.e.\ the minimum of $F_4(\vphi)$ occurs
at $\vphi_a=\vphi$.

By induction we conclude that $F_N(\vphi)$ has the minimum at
$\vphi_a=\vphi$ ($a=1,\cdots, N-1$).   It is easy to find the location 
of the minimum of 
$\tilde F_N(\vphi) = F_N(\vphi, \cdots,\vphi;\theta)$.
From the symmetry
$\vphi_N = \theta- (N-1)\vphi = \vphi$ ($mod ~2\pi$), or
$\vphi = {\theta/ N}$ ($mod ~ {2\pi/ N})$.   
Direct  evaluation of $\tilde F_N(\vphi)$ shows that
the minimum of $F_N(\vphi)$ is attained at
$\vphi_a = \bar\theta/ N$.

\bigskip

\leftline{\bf (2) Boson masses and condensates}
In a few limiting cases boson  masses and chiral condensates can be 
determined analytically.  In this subsection we suppose that fermion
masses are small $m\ll \mu$ and analyze (\ref{sym-Schro2}).  The wave
function $f(\vphi)$ is determined by two parameters, $\theta$ and
$\kappa$.   The chiral condensates are related to the boson mass by
\beqn
\la \psibar_a\psi_a \ra_\theta 
&=& \la \psibar_a\psi_a \ra_\theta' 
  - \la \psibar_a\psi_a \ra_\free' \cr
\noalign{\kern 10pt}
\la \psibar_a\psi_a \ra_\theta'  &=& 
   - {\mu_2^2\over 4\pi m} ~~~. 
\label{sym-condensate}
\eeqn

\bigskip

\noindent (a) $N=2$

We suppose that $\theta\not= \pi$.  Eq.\ (\ref{sym-Schro2}) becomes
\beqn
&&\Bigg\{ -{d^2\over d\vphi^2} -  \kappa_0 
\cos \bigg(\vphi - {\bar\theta\over 2} \bigg) \Bigg\} \, f(\vphi)
 = \ep \, f(\vphi)  \crn
&&\kappa_0 = 2 \kappa \, \cos\onehalf\bar\theta 
 =  {4mL\over \pi} \, B(\mu_1 L)^{1/2} B(\mu_2L)^{1/2}
   e^{-\pi/2\mu L} \, \cos {\bar\theta\over 2}  ~~~.
\label{sym-Schro3}
\eeqn
It is easy to see
\beeq
 \lla \cos  \vphi \rra_f =\cases{
\kappa_0 \cos \onehalf\bar\theta = \kappa(1+\cos\bar\theta) 
    &for $\kappa_0 \ll 1$\cr
\noalign{\kern 8pt}
       \cos\onehalf\bar\theta 
              &for $\kappa_0 \gg 1$~.\cr}   
\label{N=2cos}
\eneq
and accordingly
$\mu_2 L/\sqrt{2} \pi= \kappa_0$ or $\sqrt{\kappa_0}$ for $\kappa_0 \ll
1$ or $\kappa_0\gg 1$, respectively.   Hence one finds for $m \ll \mu$
\beqn
\mu_2 = \cases{
4\sqrt{2}\,  m \cos \onehalf \bar\theta  \, e^{-\pi/2\mu L} 
     &for $\mu L \ll 1$\cr 
\noalign{\kern 12pt}
4\sqrt{2} \,m \cos \onehalf \bar\theta  \,
  \bigg({\mybig e^\gamma \mu L \over\mybig 4\pi} \bigg)^{1/ 2}
 &for $\mu L \gg 1 \gg 
 (m \cos\onehalf \bar\theta)^{2/3} \mu^{1/3} L  $\cr 
\noalign{\kern 5pt}
\bigg(2 e^{\gamma} \, m\mu^{1/2} \cos \onehalf \bar\theta  \bigg)^{2/3}
    &for $ (m \cos\onehalf \bar\theta)^{2/3} \mu^{1/3} L  \gg 1$\cr}
\label{sym-mass2}
\eeqn
Coleman obtained $\mu_2 \propto  m^{2/3} \mu^{1/3} \cos^{2/3}
\onehalf \bar\theta$   in Minkowski space-time long time ago,\cite{Coleman1}
but the overall coefficient was not determined.
Note that  if the massless limit is taken with a fixed $L$, then
$\mu_2={\rm O}(m)$.

At $\theta=0$ and $L \go \infty$, $m$ dependence of $\mu_2$ has been
determined in the lattice gauge theory.\cite{Stergios} In this limit
the formula (\ref{sym-mass2}) leads to
$\mu_2/e = 2^{5/6}e_E^{2\gamma/3} \pi^{-1/6} (m/e)^{2/3}
=2.163 (m/e)^{2/3}$ where $e$ and $e_E$ are the coupling constant and
Euler's constant, respectively.  
Smilga showed that the exact coefficient should be
$2.008$ for $m/e \ll 1$.\cite{Smilga3}  The lattice simulation is done
for $m/e < .5$.  The data supports the $m^{2/3}$ dependence and indicates 
a coefficient between the two numbers mentioned above.

\bigskip

\noindent (b) $N \ge 3$ flavor

When $\kappa \gg 1$ in Eq. (\ref{sym-Schro2}), the wave function has
a sharp peak at the global minimum $\vphi_a=\bar\theta/N$, and therefore
$\lla e^{i\vphi_a} \rra_f = e^{i\bar\theta/N}$.   

For $\kappa \ll 1$ we solve (\ref{sym-Schro2}) in a power series in
$\kappa$.  To O(1) a plane wave solution
$u(\vphi ; \vec n) =  e^{in_1\vphi_1 + \cdots + in_{N-1}\vphi_{N-1}}$
satisfies $- \LapN \, u
 = \ep \, u$ where $\ep$ is positive semi-definite and vanishes only
if $n_1=\cdots=n_{N-1}=0$.  Hence to O($\kappa^0$) $f^{(0)} = 1$.  
To find O($\kappa$) correction, we note
$\LapN \, F_N(\vphi) = - F_N(\vphi)$, from which it follows that  
$f =  1 - \kappa F_N(\vphi)$.
Hence
\beeq
\lla \cos\vphi_a \rra_f = \cases{
\kappa &for $\kappa \ll 1$\cr
\noalign{\kern 8pt}
\cos \myfrac{\bar\theta}{N} &for $\kappa \gg 1$.\cr}
\label{Ncos}
\eneq
Notice that it is independent of $\theta$ for $\kappa \ll 1$.  It
follows that
$[N/8\pi^2(N-1)]^{1/2} \mu_2L = \kappa$ for $\kappa \ll 1$ and
$= \big[ \kappa \cos (\bar\theta/N) \big]^{1/2}$ for $\kappa \gg 1$.
The boson mass is determined from (\ref{sym-mass1});
\beqn
\mu_2 = \cases{
 \bigg( \myfrac{8N}{N-1} \bigg)^{1/2} m \,  e^{-\pi/N\mu L} 
       &for $\mu L \ll 1$\cr 
\noalign{\kern 10pt}
 \bigg( \myfrac{8N}{N-1}  \bigg)^{1/2} \,m \,
  \bigg( \myfrac{e^\gamma \mu L}{4\pi}  \bigg)^{1/N}
 &for $\mu L \gg 1 \gg 
 m^{N/(N+1)} \mu^{1/(N+1)} L $\cr 
\noalign{\kern 10pt}
\bigg(2 e^{\gamma} \, m\mu^{1/N} \cos \myfrac{\bar\theta}{N} 
    \bigg)^{N/(N+1)}
    &for $ m^{N/(N+1)} \mu^{1/(N+1)} L   \gg 1$\cr}
\label{sym-massN}
\eeqn

\vskip .5cm 

\sxn{General fermion masses}

\noindent (1) $N=2$

It is of interest to know the boson masses when the fermion masses 
are not degenerate.
In the two flavor case an analytic expression is obtained for
$m_a \ll \mu$ in the $L\go\infty$ limit.  

Start with (\ref{Schro4}), or more conveniently the truncated equation
(\ref{Schro6}):
\beeq
\Bigg\{ - {d^2\over d\vphi^2} - {2L\over \pi} e^{-\pi/2\mu L} 
\sum_{a=1}^2 m_a \Bbar_a \cos\vphi_a  \Bigg\}
 \, f(\vphi) = \ep_0 f(\vphi) ~.
\label{Schro10}
\eneq
Note that  $\vphi_1=\vphi$ and $\vphi_2=\theta - \vphi$.  In terms of 
$R_a=8\pi m_a \Bbar_a L^{-1} e^{\pi/2\mu L} \lla \cos\vphi_a \rra_f$,
\beqn
&&\pmatrix{\mu_1^2\cr\mu_2^2\cr} 
= \pmatrix{\mu^2\cr 0\cr} + {R_1+R_2\over 2} \crn
&&\pmatrix{\Bbar_1\cr\Bbar_2\cr}
= B(\mu_1L)^{1/2} B(\mu_2L)^{1/2} 
\left[ {B(\mu_1L)\over B(\mu_2L)} \right]^{\pm(R_1-R_2)/2\mu^2} ~~.
\label{N=2bosonmass1}
\eeqn

In the $L\go\infty$ limit, $\Bbar_a \propto L$ so that 
$\lla \cos \vphi \rra_f = \cos \vphi_{\rm min}$ where the potential term
in (\ref{Schro10}) has a minimum at $\vphi_{\rm min}$.  In general
at the minimum of  a function $g(\vphi)= -\alpha\cos \vphi -
\beta\sin\vphi$, $e^{i\vphi}=(\alpha+i\beta)/\sqrt{\alpha^2+\beta^2}$.
Hence,
\beeq
R_a = 8\pi \,
{(m_a\tB_a)^2 + m_1m_2\tB_1\tB_2\cos\theta \over
\sqrt{(m_1\tB_1)^2 + (m_2\tB_2)^2 + 2m_1m_2\tB_1\tB_2\cos\theta} }
\label{N=2R1}
\eneq
where $\tB_a = \Bbar_a/L$.  Solving (\ref{N=2bosonmass1}) and 
(\ref{N=2R1}) to the leading order in $m_a/\mu$, one finds
\beeq
R_a = 2 e^{4\gamma/3} \mu^{2/3}
{m_a^2 + m_1 m_2 \cos\theta \over
(m_1^2 + m_2^2 + 2m_1m_2\cos\theta)^{1/3} } ~.
\label{N=2R2}
\eneq
Consequently, 
\beeq
\mu_2 = e^{2\gamma/3} \mu^{1/3}
     (m_1^2 + m_2^2 + 2m_1m_2\cos\theta)^{1/3}   
\label{N=2bosonmass2}
\eneq
and
\beeq
\pmatrix{ \la\psibar_1 \psi_1 \ra_\theta' \cr
          \la\psibar_2 \psi_2 \ra_\theta' \cr} =
- {e^{4\gamma/3} \mu^{2/3} \over 2\pi 
(m_1^2 + m_2^2 + 2m_1m_2\cos\theta)^{1/3} } 
\pmatrix{m_1+m_2\cos\theta\cr m_2+m_1\cos\theta \cr} ~.
\label{N=2condensate1}
\eneq
In the symmetric case $m_1=m_2=m$ (\ref{N=2bosonmass2}) reduces to
(\ref{sym-mass2}).  

Observe that there is no singularity at $\theta=\pi$ in a generic case
$m_1 \not= m_2$.  The boson mass and chiral condensates are smooth
functions of $\theta$ with a period $2\pi$.  The singularity appears only
when $m_1=m_2\not=0$ in the two flavor case.   
In a special case $m_1=0$ but $m_2\not=0$ ($m_2\ll\mu$), 
$(\la\psibar_1 \psi_1 \ra_\theta',\la\psibar_2 \psi_2 \ra_\theta')$=$
- (2\pi)^{-1} (e^{4\gamma} \mu^2 m_2 )^{1/3} (\cos\theta, 1)$ in this
approximation.

\bigskip

\noindent (2) $N=3$

The three flavor case mimics physics of four-dimensional QCD. 
Three fermions, which one may call ``up'', ``down'', and ``strange''
quarks, have different masses; $m_1 \sim m_2 \ll m_3$.  We would like to
see how the asymmetry in masses affects chiral condensates and boson
masses.  

\begin{figure}[bt]
\vskip 0.3cm
\hskip 4cm
\epsfxsize= 8.cm
\epsffile[75 310 391 602]{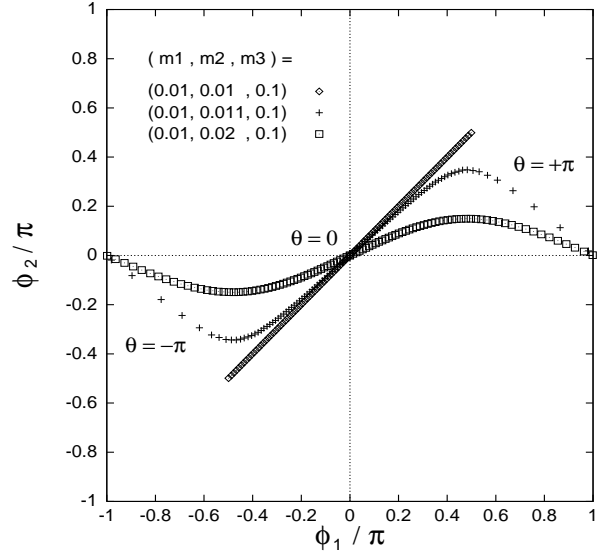}
\vskip -0cm 
\caption{\small The location of the global minimum of the potential (9.7)
in the $N=3$ model with general fermion masses.  As $\theta$ changes
from $-\pi$ to $\pi$, the minimum in the figure moves from left to right.
The values for the fermion masses, $m_a$'s, are in a unit of $\mu$.  The
location of the minimum discontinuously jumps at $\theta=\pm \pi$ for
$(m_1,m_2,m_3)/\mu=(0.01, 0.01, 0.1)$ or $(0.01, 0.011, 0.1)$, but
makes a continuous loop for $(0.01, 0.02, 0.1)$.}
\label{fig.5}
\end{figure}

We concentrate on the $L\go\infty$ ($T=0$) limit in the truncated
theory.  As in the $N=2$ case one needs to find the location of the 
minimum of the potential
\beeq
g(\vphi_1, \vphi_2;\theta) 
= \lim_{L\go\infty} {\pi\over 3L^2} \, V(\vphi) 
= - \sum_{a=1}^3 m_a \tB_a \cos \vphi_a ~~.
\label{N=3pot1}
\eneq
where $\vphi_3 \equiv \theta - \vphi_1 - \vphi_2$.
Here $\tB_a$'s  are determined by the boson masses $\mu_\alpha$'s and
eigenvectors:
\beqn
\tB_a &=&
{e^\gamma\over 4\pi} \prod_{\alpha=1}^3 \mu_\alpha^{(U_{\alpha a})^2}\cr
\noalign{\kern 10pt}
{\cal K} \, &=&
{\mu^2\over 3} \pmatrix{1&1&1\cr 1&1&1\cr 1&1&1\cr} +
\pmatrix{R_1 \cr &R_2\cr &&R_3\cr} 
= U^t \pmatrix{\mu_1^2\cr &\mu_2^2 \cr &&\mu_3^2 \cr} U \cr
\noalign{\kern 10pt}
R_a &=& 8\pi m_a \tB_a \cos \vphi_a^{\rm min} ~.
\label{N=3matrix1}
\eeqn
$\vphi_a^{\rm min}$ is the location of the minimum of $g(\vphi)$. 
The set of equations (\ref{N=3pot1}) and
(\ref{N=3matrix1}) must be solved simultaneously.  Chiral 
condensates are given by
\beeq
\la \psibar_a \psi_a \ra_\theta = - 2 \tB_a \cos \vphi_a^{\rm min}
+ {e^{2\gamma}\over \pi} m_a ~.
\label{N=3condensate1}
\eneq

When fermion masses are degenerate,  
$\vphi_a^{\rm min}= {1\over 3} \bar\theta$, as shown in Section 9.  At
$\theta=\pi$ the location of the minimum changes discontinuously, which
induces singular behavior in physical quantities.  We show that the
singularity disappears if the asymmetry in fermion masses is 
sufficiently large.
In ref.\ \cite{HHI2} this problem was analyzed by examining $g(\vphi)$,
but without solving (\ref{N=3matrix1}).

When $m_1=m_2<m_3$, $\vphi_1=\vphi_2$ at the minimum of $g(\vphi)$.
At $\theta=0$, $\vphi_1=\vphi_2=0$.  As $\theta$ increases,
$\vphi_1=\vphi_2$ also increases.  At $\theta=\pi$ it reaches
$\vphi_1=\vphi_2=\vphi_c$ whose value depends on $m_a$'s.  As
$m_3$ gets bigger and bigger with $m_1=m_2$ kept fixed, $\vphi_c$
approaches $\onehalf \pi$.  For instance $\vphi_c=(0.467, 0.486,
0.499)\, \pi$ for $m_1/\mu=0.01$ and $m_3/\mu=(0.02, 0.03, 0.1)$.
As $\theta$ exceeds $\pi$, the minimum jumps to
$\vphi_1=\vphi_2=-\vphi_c$ and returns to $\vphi_1=\vphi_2=0$ at
$\theta=2\pi$.  The singular behavior at $\theta=\pi$ remains.
This is expected as $m_3\gg m_1=m_2$ corresponds to the two flavor
case.

\begin{figure}[bt]
\vskip 0.3cm
\hskip 3cm
\epsfxsize= 9.5cm
\epsffile[45 290 473 601]{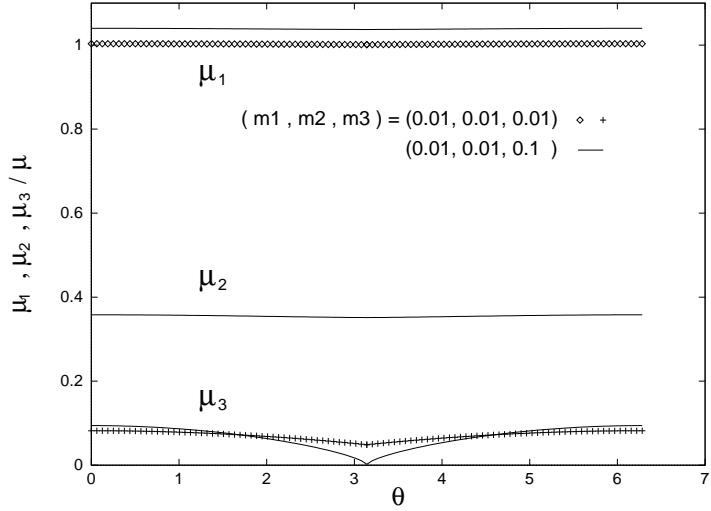}
\vskip -0cm 
\caption{\small Boson masses in the $N=3$ model at $T=0$
with degenerate fermions
$m_1=m_2=m_3$ and with $m_1=m_2<m_3$. In the former case $\mu_2=\mu_3$.
Fermion masses are in a unit of $\mu$.}
\label{fig.6}
\end{figure}

Now we add a small asymmetry in the light fermions.  Fig.\ 5
depicts the location of the minimum of $g(\vphi)$ as $\theta$
varies from $-\pi$ to $\pi$. A small asymmetry in $m_1$ and $m_2$,
does not change the behavior near $\theta=0$, but significantly affects
the behavior near $\theta = \pm \pi$.  
At $(m_1,m_2,m_3)/\mu = (0.01, 0.011, 0.1)$  the minimum at $\theta = \pi$
is very close, but not quite equal, to ($\pi, 0$). 
At $(m_1,m_2,m_3)/\mu = (0.01, 0.02, 0.1)$ the minimum at $\theta=\pi$ is
located at ($\pi, 0$), and the singularity in physical quantities at
$\theta=\pi$ disappears.

In fig.\ \ref{fig.6} we have plotted boson masses $\mu_1$, $\mu_2$,
and $\mu_3$ as functions of $\theta$ with given fermion masses.
They correspond to $m_{\eta '}$, $m_\eta$, $m_\pi$ in QCD.
For $m_1=m_2=m_3\ll \mu$, $\mu_1 \gg \mu_2=\mu_3$.  When 
$m_1=m_2< m_3\ll \mu$, all $\mu_\alpha$'s are different.
The $\theta$ dependence of each $\mu_\alpha$ has similar behavior.
$\mu_1/\mu$,  $\mu_2/\mu$, and $\mu_3/\mu$  vary by 0.003, 0.007, and
0.1 in magnitude.  The mass of the lightest boson, $m_\pi$, 
has the most $\theta$  dependence.

\begin{figure}[bt]
\vskip 0.3cm
\hskip 3cm
\epsfxsize= 9.5cm
\epsffile[45 290 473 603]{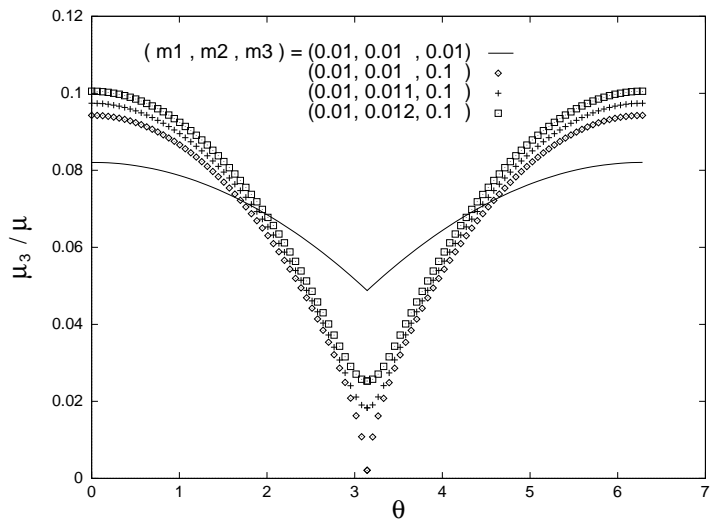}
\vskip -0cm 
\caption{\small The mass, $\mu_3$,  of the lightest boson in the $N=3$
model at $T=0$ for various values of fermion masses. Fermion masses are in
a unit of $\mu$.  The cusp at $\theta=\pi$ disappears as a small asymmetry
in the two light fermions is added.}
\label{fig.7}
\end{figure}

In fig.\ \ref{fig.7} the $\theta$ dependence of the mass of the lightest
boson is plotted for various values of fermion masses.  The cusp at
$\theta=\pi$ persists so long as $m_1=m_2$, but a small asymmetry in
$m_1$ and $m_2$ changes it to  smooth dependence.  
The mass $\mu_3$ at $\theta=0$ increases as the third fermion 
gets heavier as expected, whereas  it decreases at $\theta=\pi$.

\begin{figure}[bt]
\vskip 0.3cm
\hskip 3cm
\epsfxsize= 9.5cm
\epsffile[29 290 473 601]{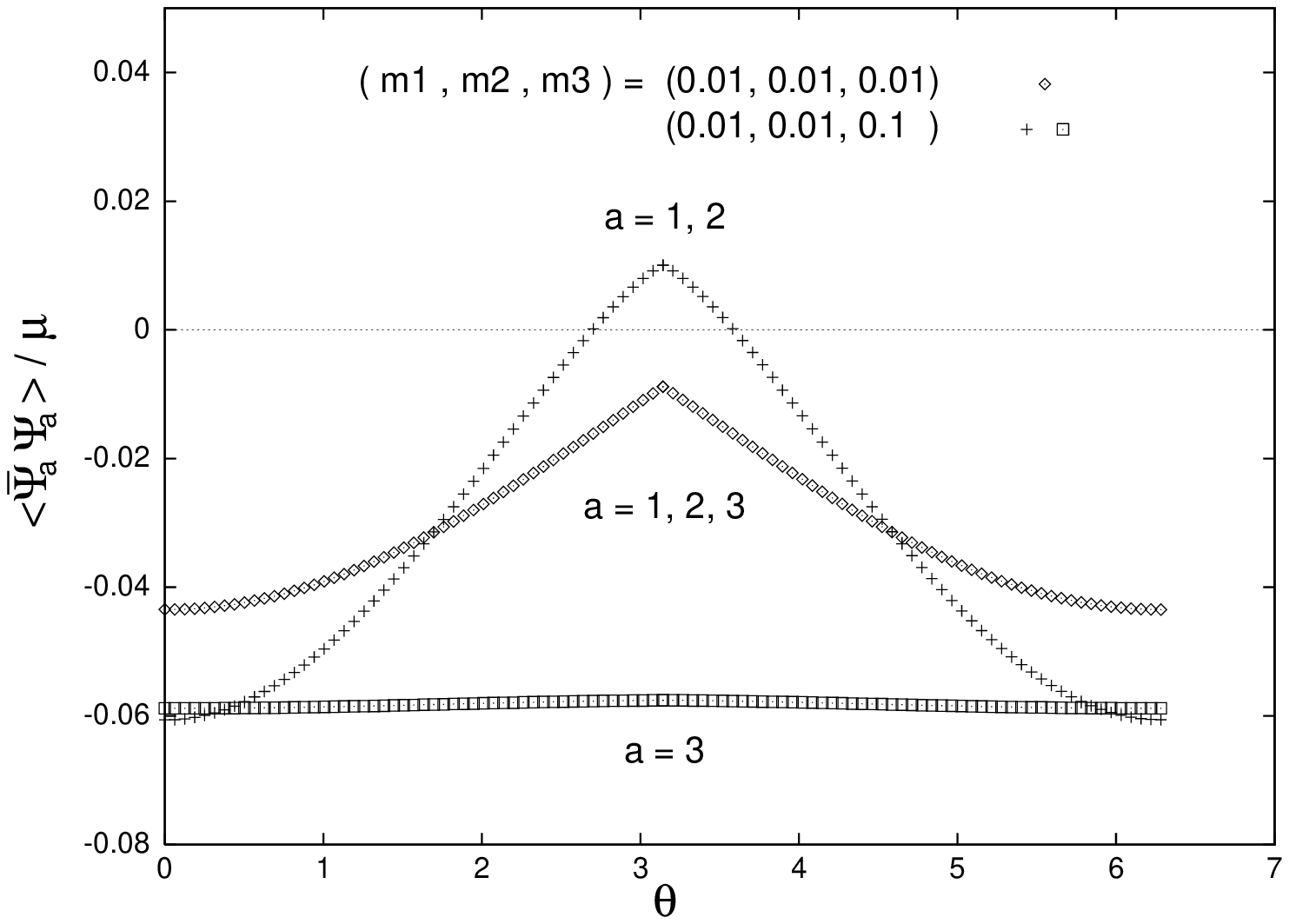}
\vskip -0cm 
\caption{\small The $\theta$ dependence of the chiral condensates in the $N=3$ 
model at $T=0$. Two cases, $(m_1,m_2,m_3)/\mu = (.01, .01, .01)$ and
$(.01, .01, .1)$, are displayed. In the latter case the condensate of
the heavy fermion has little dependence on $\theta$, whereas the 
condensates of the light fermions show large dependence.}
\label{fig.8}
\end{figure}

The $\theta$ dependence of the location of the minimum of $g(\vphi)$ and
the value of the mass $\mu_3$ of the lightest boson induces 
nontrivial $\theta$ dependence in the chiral condensates
$\la\psibar_a\psi_a\ra_\theta$.  In fig.\ \ref{fig.8} a chiral
condensates for $(m_1,m_2,m_3)/\mu=(0.01,0.01,0.01)$ and
(0.01,0.01,0.1) are depicted.  In both cases there appear  
cusps at $\theta=\pi$.  Notice that the magnitude of the condensates
at $\theta=0$ is insensitive to fermion masses.  This
is true only if the ``free theory background'', $\la\psibar\psi\ra_\free'$ is
subtracted in the  definition of the condensate (\ref{N=3condensate1}).  

There appears, however, a big difference in the $\theta$ dependence
of the condensates. 
When $(m_1,m_2,m_3)/\mu=(0.01,0.01,0.1)$, the third fermion $\psi_3$
is much heavier than the other two.  The condensate 
$\la\psibar_3\psi_3\ra_\theta$ is more or less independent of $\theta$,
which is expected as the vacuum structure is mainly determined by light
fermions.

\begin{figure}[bt]
\vskip 0.3cm
\hskip 3cm
\epsfxsize= 9.5cm
\epsffile[29 290 473 601]{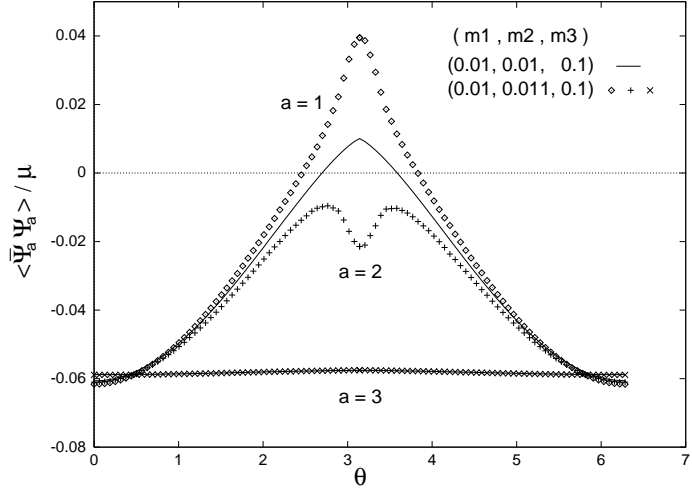}
\vskip -0cm 
\caption{\small The $\theta$ dependence of the chiral condensates in the $N=3$ 
model at $T=0$. Small asymmetry is added to masses of light fermions.
The effect is minor near $\theta=0$, but the condensates of the light fermions 
near $\theta=\pi$ are significantly affected.}
\label{fig.9}
\vskip 0.3cm
\end{figure}

When a small asymmetry in light fermions is added, condensates suffer
a big change.  See fig.\ \ref{fig.9}.  With
$(m_1,m_2,m_3)/\mu=(0.01,0.011,0.1)$ the $\theta$ dependence in
$\la\psibar_1\psi_1\ra_\theta$ is enhanced, whereas 
$\la\psibar_2\psi_2\ra_\theta$ develops a dip near $\theta=\pi$.
A small asymmetry in $m_1$ and $m_2$ induces a big difference in
$\la\psibar_1\psi_1\ra_\theta$ and $\la\psibar_2\psi_2\ra_\theta$
near $\theta=\pi$.   The nonlinearity in Eq.\ 
(\ref{N=3pot1}) and (\ref{N=3matrix1}) gives rise to such sensitive 
dependence.

It is interesting to recognize the similarity between the potential
$g(\vphi)$ in (\ref{N=3pot1}) and the effective chiral
Lagrangian proposed by Witten to describe low energy behavior of
four-dimensional QCD.\cite{Witten}  In  Witten's approach
\beeq 
V^{\rm Witten}(U) = 
f_\pi^2 \bigg\{ - {1\over 2} {\rm Tr}\, M(U+U^\dagger) + 
{k\over 2N_c} (-i \ln \det U - \theta)^2 \bigg\}
\label{Witten1}
\eneq
where $U$ is the pseudo-scalar field matrix and $M= diag (m_u,m_d,m_s)$ is
the quark mass matrix.  The second term represents contributions from
instantons.  $k$ is O(1) in the large $N_c$ (color) limit.

The fact that $m_{\eta'}^2 \gg m_\pi^2, m_K^2, m_\eta^2$ implies that 
$k/N_c \gg m_a$.  Diagonalize $U$ and denote it by $diag(e^{i\phi_1},
e^{i\phi_2}, e^{i\phi_3})$.  As the second term dominates over the 
first, $\sum \phi_a = \theta$ to the first approximation.  With
$\phi_3$ eliminated, $V^{\rm Witten}(U)$ takes the same form as 
$g(\vphi)$ in (\ref{N=3pot1}).  Consequently both models show
qualitatively similar behavior.  Indeed, Witten has argued that
a small asymmetry in $m_u$ and $m_d$, in addition to large asymmetry
$m_s \gg m_u, m_d$ removes the singularity of physical quantities 
in $\theta$ at $\pi$, which is exactly what we are observing in
the $N=3$ Schwinger model.  However, it should be noted that the
coefficients in the  potential $g(\vphi)$ have extra fermion mass
dependence coming from the factors $\tB_a$, which have significant
effects near 
$\theta=\pi$.

\vskip .5cm

\sxn{Summary}

In this paper the massive N-flavor Schwinger model was analyzed 
in the generalized Hartree-Fock approximation. Dynamics of the 
zero-modes is determined by the Schr\"odinger equation for N degrees of
freedom.  The potential term in the Schr\"odinger equation depends
on the boson spectrum in the oscillatory modes.  The boson mass spectrum
in turn depends on the ground state in the zero-mode sector.  The ground
state of the two sectors must be determined self-consistently.  

We have evaluated the boson spectrum and chiral condensates in
the $N=1$, 2, 3 models.  In the $N=1$ model we have found anomalous
dependence of physical quantities on the fermion mass near $\theta=\pi$
at low temperature.  In the $N=3$ model physics near $\theta=\pi$ is
very sensitive to the small asymmetry in fermion masses.  Chiral
dynamics in the $N=3$ model resembles with that in QCD in
four dimensions.

\vskip 1cm

{\small \baselineskip=10pt  
\leftline{\bf Acknowledgment}
The authors would like thank to Philippe  de Forcrand and  Jim  Hetrick 
for useful discussions and communication, 
and Konstantinos Stergios for enlightening communication.
This work was supported in part 
by  the U.S.\ Department of Energy under contracts DE-FG02-94ER-40823.
}

\vskip 1cm 
\axn{Correlation functions}

Green's function for a scalar field with a mass $\mu$ on $S^1$,
excluding the zero mode, is
\beeq
\Delta (x;\mu , L) = \sum_{n=1}^\infty
{1\over \sqrt{(2\pi n)^2 + (\mu L)^2 } } ~ \cos {2\pi inx\over L}  ~.
\label{GreenF1}
\eneq
In the massless case ($\mu=0$) it is given by
\beqn
\Delta (x;0 , L) 
&=& - {1\over 4\pi} \ln 2\Big( 1- \cos{2\pi x\over L} \Big) \cr
\noalign{\kern 6pt}
e^{2\pi \Delta (x;0 , L)} &\sim& {L\over 2\pi x} 
\qquad
\hbox{for } \Big| {2\pi x\over L} \Big| \ll 1 ~~. 
\label{GreenF2}
\eeqn
In terms of
\beqn
I[\, s; a,b\, ] = 
 \sum_{n=1}^\infty  {\cos 2\pi nb \over (n^2 + a^2)^s }   ~,
\label{Ifunction1}
\eeqn
$2 \pi \Delta (x;\mu , L) = I[\, \onehalf ; \mu L/2\pi,x/ L\, ]$ and
$B(\mu L) = \exp \Big\{ - I[\, \onehalf ; \mu L/2\pi, 0\,] 
    + I[\,\onehalf ; 0, 0\, ] \Big\}$.  
For $|b|<1$ 
\beqn
I[\, s; a,b\, ] &=& \int_0^\infty dt \, {\cos 2\pi b t\over (t^2+a^2)^s}
- {1\over 2 a^{2s}} + 2 \sin s\pi \int_a^\infty dt \,
{\cosh 2\pi bt\over (t^2-a^2)^s (e^{2\pi t} -1 )}  \cr
\noalign{\kern 10pt}
&=& {\sqrt{\pi}\over \Gamma(s)} \, \Big| {\pi b\over a} \Big|^{s-\onehalf} \,
K_{s-\onehalf}(2\pi|ab|)  - {1\over 2 a^{2s}} \cr
\noalign{\kern 10pt}
&&\hskip 3cm 
+ 2 \sin s\pi \int_a^\infty dt \,
{\cosh 2\pi bt\over (t^2-a^2)^s ( e^{2\pi t} -1)} ~.
\label{Ifunction2}
\eeqn
The above formula is valid for an arbitrary $s$ by analytic continuation.

It then follows that 
\beqn
&&2 \pi \Delta (x;\mu , L) = 
 K_0(|\mu x|)   - {\pi\over \mu L} + 2  \int_1^\infty du \,
{\cosh \mu xu\over  ( e^{\mu L u}-1 ) \sqrt{u^2-1}} ~~. 
\label{GreenF4}
\eeqn
Recalling that 
$K_0(z) \sim - \ln z - \gamma + \ln 2$ for $z \ll 1$ and 
$\sim \sqrt{{\pi\over 2z}} \, e^{-z}$ for  $z \gg 1$,  
and noticing for $\mu L \gg 1$
\beeq
 2  \int_1^\infty du \, 
{\cosh \mu xu\over  ( e^{\mu L u}-1 ) \sqrt{u^2-1}} 
\sim 
 \sqrt{ {\pi\over 2\mu(L-x)} } e^{-\mu(L-x)}
 + \sqrt{ {\pi\over 2\mu(L+x)} } e^{-\mu(L+x)} ~~~,  
\nonumber
\eneq
one finds 
\beeq
e^{2\pi \Delta(x ; \mu ,L) } \sim
\cases{
 ~~ 1
&for $\mu L \gg1~,~ \myfrac{x}{L} \ll 1 ~,~ \mu x \gg 1$\cr
\noalign{\kern 6pt}
\myfrac{2 e^{-\gamma}}{\mu  x} 
&for $\mu L \gg1~,~ \myfrac{x}{L} \ll1 ~,~ \mu x \ll 1$~.\cr}
\label{GreenF5}
\eneq

\vskip 1cm


\axn{Normal ordering and Bogoliubov transformation}

In the subsequent discussions, we make frequent use of identities: (1)
$: e^A:  = e^{A^-} e^{A^+}$, (2) 
$ e^A \, e^B = e^{{1\over 2} [A,B]} \, e^{A+B} = e^{[A,B]} \, e^B \, e^A$,
and (3) $: e^A: \,  :e^B :   = e^{[A^+ , B^-]}   :e^{A+B}:$ 
~where $A^+$ and $A^-$ denote the annihilation and creation operator parts
of $A$, respectively.  With massless fields (\ref{BoseVariables1})
\beeq
[~\phi^a_\pm (t,x)^{(+)} ~,~ \phi^b_\pm (0,0)^{(-)} ~]  
 = - \delta^{ab} \, {1\over 4\pi} \, 
\ln \big\{ 1 - e^{-2\pi i (t \pm x - i\ep)/L} \big\} ~~.
\label{formula2}
\eneq
We also note that
\beqn
&&{1\over L} \, \left\{ 
{e^{+i \pi x/L}\over 1 -  e^{+2\pi i (x+i\ep)/L}} +
{e^{-i \pi x/L}\over 1 -  e^{-2\pi i (x-i\ep)/L}}  \right\}
  = e^{i\pi x/L} \, \delta_L (x)   \cr
\noalign{\kern 4pt}
&&{e^{+i \pi x/L}\over 1 -  e^{+2\pi i (x+i\ep)/L}} =
{-1\over 2\pi i} ~ \Big( {L\over x+ i\ep} + {\pi^2\over 6} {x\over L} 
 + {7\pi^4\over 360} {x^3\over L^3} + \cdots \Big)  ~. 
\label{formula3}
\eeqn

We have seen that boson fields become massive due to the Coulomb
interaction and fermion masses.  When this happens,
the vacuum also changes and in two dimensions non-vanishing
chiral condensates result. 

It is most convenient to work in the Schr\"odinger
picture.   A boson field is generically denoted by $\phi(x)$ with
its conjugate  $\Pi(x)$. On a circle  they are expanded as
\beqn
\phi(x)   &=& \sum_{n\not=0}{1\over \sqrt{2\omega_n(\mu) L} }
\Big\{ c_{n}^{}(\mu) e^{ ip_nx} + c_{n}^\dagger(\mu) e^{-ip_nx}
\Big\} \crn
\Pi(x)  &=& - i \sum_{n\not=0}\sqrt{{\omega_n(\mu)\over 2L}}
\Big\{ c_{n}^{}(\mu) e^{ ip_nx} - c_{n}^\dagger(\mu) e^{-ip_nx}
\Big\} 
\label{phi-expansion1}
\eeqn
where $p_n= {2\pi n/ L}$ and $\omega_n(\mu) = \sqrt{p_n^2 + \mu^2}$.
Annihilation and creation operators $c_n(\mu)$ and
$c_n^\dagger(\mu)$ are defined with respect to a mass $\mu$.
The left- and right-moving modes, $\phi_+(x)$ and $\phi_-(x)$, of
$\phi=\phi_+ + \phi_-$ are defined by the $n<0$ and $n>0$ components
in (\ref{phi-expansion1}).
In the massless limit they corresponds to (\ref{BoseVariables1}) in
the  Schr\"odinger picture.  

Sometimes we need to treat  $\phi_{\pm}$'s
separately. One finds
\beqn
&&\phi_{\pm} (x) = {1\over 2} \phi(x) \pm 
{1\over 2L} \int_0^L dy \, F(x-y; \mu) \, \Pi_\phi(y) \crn
&&F(x;\mu) = \sum_{n=1}^\infty {i\over \omega_n(\mu)} \Big\{ 
e^{-2\pi inx/L} - e^{+2\pi inx/L} \Big\} ~~.
\label{phi-identity1}
\eeqn
Note that the definition of $\phi_\pm(x)$ depends on the reference mass
$\mu$ as opposed to that of $\phi(x)$ and $\Pi_\phi(x)$.  In the massless
theory $F'(x; 0) = L \delta_L(x) - 1$ so that 
$\Pi(x) =  \phi_{+}'(x) - \phi_{-}'(x) $.

We have an identity among $c_n(\mu)$'s with different $\mu$'s:
\beqn
&&c_n(\mu_1) = \cosh \theta_n(\mu_1;\mu_2) c_n^{}(\mu_2) 
            + \sinh \theta_n(\mu_1;\mu_2) c_{-n}^\dagger(\mu_2) \crn
&&\left[ \matrix{ \cosh \theta_n(\mu_1;\mu_2) \cr
                  \sinh \theta_n(\mu_1;\mu_2) \cr} \right] 
= {1\over 2} \left\{ \sqrt{ {\omega_n(\mu_1)\over \omega_n(\mu_2)}
}
\pm \sqrt{ {\omega_n(\mu_2)\over \omega_n(\mu_1)} } \right\} ~~.
\label{cc-identiry1}
\eeqn
In other words the change in the boson mass induces a Bogoliubov
transformation.  The vacuum with respect to a boson mass $\mu$
is defined by $c_n(\mu) ~ | {\rm vac} ; \mu  \ra = 0 $.

In our formalism fermion fields are first bosonized in the interaction
picture defined by massless bosons.  Boson fields then acquire masses
and the vacuum is redefined.  In particular  boson fields are 
normal-ordered with respect to physical boson masses.  One useful relation is 
\beeq
e^{i\alpha\phi(x)} = \exp \Bigg\{ - {\alpha^2\over 2L}
 \sum_{n=1}^\infty {1\over \omega_n(\mu)} \Bigg\} ~
N_\mu[e^{i\alpha\phi(x)}] ~,
\label{normal-order1}
\eneq
from which it follows \cite{Coleman2,HH}
\beqn
N_0[e^{i\alpha\phi(x)}] &=& B(\mu L)^{\alpha^2/4\pi} 
~ N_\mu[e^{i\alpha\phi(x)}] \crn
B(\mu L) 
&=& \exp \Bigg\{ - \sum_{n=1}^\infty 
\Big( {1\over \sqrt{ n^2 + (\mu L/2\pi)^2} } - {1\over n} \Big) \Bigg\} \crn
&=& {\mu L\over 4\pi} \exp \bigg\{ \gamma + {\pi\over \mu L}
 - 2 \int_0^\infty  {dx \over e^{ \mu L \cosh x} - 1}  \bigg\} ~.
\label{B-function1}
\eeqn

In the following we make use of simplified notation:  $\phi_\pm(x)$ refers
to the massless field, $\omega_n = \omega_n(0)$ and
$\theta_n=\theta_n(0;\mu)$.  $N_\mu[A(c, c^\dagger)]$ denotes that the
operator $A$ is normal-ordered with respect to $c_n(\mu)$ and
$c_n^\dagger(\mu)$.  Some useful identities are
\beeq
N_0[\, e^{i\alpha\phi(x) + i\beta\phi(y)} \,] 
\Big/  N_\mu [\, e^{i\alpha\phi(x) + i\beta\phi(y)} \,]
     = B(\mu L)^{(\alpha^2+\beta^2)/4\pi} \, 
e^{-\alpha\beta \{ \Delta(x-y;\mu,L) - \Delta(x-y;0,L) \} } 
\label{normal-order2}
\eneq
\beqn
&&N_0[e^{i\alpha_+ \phi_+(x) + i\alpha_- \phi_-(x)}] 
\Big/ N_\mu[e^{i\alpha_+ \hat\phi_+(x) 
       + i\alpha_- \hat\phi_-(x)}]  \crn
&&\hskip 3cm =\exp \bigg\{ - \sum_{n=1}^\infty {1\over 2\omega_nL}
\Big[ 2\alpha_+\alpha_- \cosh\theta_n \sinh\theta_n 
+ (\alpha_+^2 + \alpha_-^2) \sinh^2 \theta_n \Big] \bigg\} \crn
&&\hskip 1cm \hat\phi_\pm(x;\mu) = \sum_{n=1}^\infty
{1\over \sqrt{2\omega_n L}} \bigg\{
\pmatrix{\sinh\theta_n \cr \cosh\theta_n \cr} 
  (c_n(\mu) e^{ip_nx} + c_n^\dagger(\mu) e^{-ip_nx} ) \crn
&&\hskip 5.cm
 + \pmatrix{\cosh\theta_n \cr \sinh\theta_n \cr} 
  (c_{-n}(\mu) e^{-ip_nx} + c_{-n}^\dagger(\mu) e^{ip_nx} )  \bigg\} ~~,
\label{normal-order3}
\eeqn

\beqn
&&N_0[\, e^{i\alpha\phi_+ (x) - i\alpha\phi_- (x) 
  + i\beta\phi_+(y) - i\beta\phi_- (y) } \,] \Big/
N_\mu [\, e^{i\alpha \hat\chi_+ (x) - i\alpha\hat\chi_- (x) 
  + i\beta\hat\chi_+(y) - i\beta\hat\chi_- (y) } \,] 
\crn
&&\hskip 1cm = B(\mu L)^{(\alpha^2+\beta^2)/4\pi} \, 
e^{-\alpha\beta \{ \Delta(x-y;\mu,L) - \Delta(x-y;0,L) \} } 
\, e^{- \onehalf (\alpha^2+\beta^2)  h(0; \mu,L) 
 - \alpha\beta h(x-y; \mu ,L) } \crn
&&\hskip 2cm \hat\chi_\pm(x) = \sum_{n=1}^\infty 
{ 1 \over  \omega_n  }
\sqrt{ {\omega_n(\mu)\over 2L} } ~ \Big( 
c_{\mp n}(\mu)  \, e^{\mp ip_nx} + {\rm h.c.}   \, \Big) \qquad \cr
\noalign{\kern 10pt}
&&\hskip 2cm h(x;\mu,L) = {1\over 2L} \sum_{n\not= 0} 
{ \mu^2 \over \omega_n^2 \omega_n(\mu) } ~ e^{i p_n x} ~.
\label{normal-order6}
\eeqn

\vskip 1cm


\axn{Useful identities for zero modes}

\noindent (1) Free massless fermion

In a theory of a free massless fermion the Hamiltonian in the bosonization
method is given by 
\beeq
H = {\pi\over 2L} (4 p^2 + \tilde p^2) 
+ \int dx {1\over 2} (\dot \phi^2 + \phi'^2)  
\label{formula-f1}
\eneq
where $\phi=\phi_+ +\phi_-$ represents a massless boson defined 
in (\ref{BoseVariables1}).  The vacuum state is
$|\Psi_\vac\ra = |n=0\ra$, i.e.\  $p ~ |\Psi_\vac\ra =0$.  Note that
$\la e^{i\alpha q} \ra=\delta_{\alpha,0}$.  Since $q(t)=q+4\pi pt/L$,
\beqn
\la e^{iaq(t)} e^{ibq(0)} \ra ~~ &=& \delta_{a+b}\, e^{-2\pi i a^2 t/L}
\cr
\la T[e^{iaq(t)} e^{ibq(0)}] \ra &=&
  \delta_{a+b}\,  e^{-2\pi ia^2|t|/L} ~~~.
\label{formula-f2}
\eeqn

\bigskip

\noindent (2) QED$_2$

If all fermions are massless, the  model is solved in the operator form
in the Heisenberg picture.  The solution to (\ref{Hamiltonian5}) in the
$N$-flavor model is
\beqn
\Wil'(t) &=& \Wil' \cos \mu t + {\pi \mu L\over N} \, P_W \sin \mu t\crn
P_W(t) &=& P_W \cos \mu t - {N\over \pi\mu L} \,\Wil' \sin\mu t \crn
q_a'(t) &=& q_a' + {4\pi t\over L} \big(p_a -{1\over N} \sum p_b\big)\crn
p_a(t) &=& p_a \crn
\tq_a(t) &=& \tq_a + {\pi t\over L} \tp_a \crn
\tp_a(t) &=& \tp_a ~~.
\label{masslessSol1}
\eeqn
Here $\Wil'$ and $q_a'$ are defined in (\ref{newWil}).

In the $N=1$ model
\beeq
\la e^{i\ell q + i\beta(\Wil + 2\pi p) + i\gamma p_W} \ra
= e^{i\ell \theta}  
e^{-  \pi \mu L \beta^2/4  - (\gamma-2\pi \ell)^2 /4\pi \mu L}  ~.
\label{formula-1}
\eneq
It follows from (\ref{masslessSol1}) and (\ref{formula-1}) that
\beeq
\la e^{i\alpha q(t) } e^{i\beta q(0)} \ra
= \exp \left\{ i(\alpha +\beta)\theta 
  -{\pi(\alpha^2 +\beta^2)\over \mu L} 
- {2\pi\alpha\beta\over \mu L} e^{-i\mu t} \right\} ~~. 
\label{formula-2}
\eneq

\vskip .5cm


\def\jnl#1#2#3#4{{#1}{\bf #2} (#4) #3}

\def\ap {{\it Ann.\ Phys.\ (N.Y.)} }
\def\cmp {{\it Comm.\ Math.\ Phys.} } 
\def\ijmpA {{\it Int.\ J.\ Mod.\ Phys.} {\bf A}} 
\def\ijmpB {{\it Int.\ J.\ Mod.\ Phys.} {\bf B}} 
\def\ijmpC {{\it Int.\ J.\ Mod.\ Phys.} {\bf C}} 
\def\jpA {{\it J.\ Phys.\ {\bf A} Math.\ Gen.\ }}
\def\jpG {{\it J.\ Phys.\ {\bf G}}}
\def\jmp {{\it  J.\ Math.\ Phys.} } 
\def\mplA {{\it Mod.\ Phys.\ Lett.} {\bf A}} 
\def\mplB {{\it Mod.\ Phys.\ Lett.} {\bf B}} 
\def\plB {{\it Phys.\ Lett.} {\bf B}} 
\def\plA {{\it Phys.\ Lett.} {\bf A}} 
\def\nc {{\it Nuovo Cimento} } 
\def\npB {{\it Nucl.\ Phys.} {\bf B}} 
\def\npps{{\it Nucl.\ Phys.\ Proc.\ Suppl.\ }}
\def\pr {{\it Phys.\ Rev.} } 
\def\prl {{\it Phys.\ Rev.\ Lett.} } 
\def\prB {{\it Phys.\ Rev.} {\bf B}} 
\def\prD {{\it Phys.\ Rev.} {\bf D}} 
\def\prp {{\it Phys.\ Report} } 
\def\ptp {{\it Prog.\ Theoret.\ Phys.} } 
\def\ptps {{\it Prog.\ Theoret.\ Phys.\ Suppl.} } 
\def\rmp {{\it Rev.\ Mod.\ Phys.} } 
\def\Zp {{\it Z.\ Phys.} }
\def\hep {{\tt hep-th/}}

\myend
